\documentclass[aps,prd,reprint,groupedaddress,amsmath]{revtex4-2}

\usepackage[colorlinks,linkcolor=blue,hyperfootnotes=true]{hyperref}
\usepackage{amsmath}
\usepackage{graphicx}
\usepackage{dcolumn}
\usepackage{bm}
\usepackage{footnote}
\usepackage{extarrows}
\usepackage{longtable}
\usepackage{threeparttable}
\usepackage{tabularx}
\usepackage{adjustbox}
\usepackage{amssymb}
\allowdisplaybreaks[2]

\begin{document}


\title{The Existence and Distribution of Photon Spheres Near Spherically Symmetric \\ Black Holes --- A Geometric Analysis}


\author{Chen-Kai Qiao}
\email{Email: chenkaiqiao@cqut.edu.cn}
\affiliation{College of Science, Chongqing University of Technology, Banan, Chongqing, 400054, China}



\date{\today}

\begin{abstract}
Photon sphere has attracted significant attention since the capture of black hole shadow images by Event Horizon Telescope. Recently, a number of studies have highlighted that the number of photon spheres and their distributions near black holes are strongly constrained by black hole properties. Specifically, for black holes with event horizons and proper asymptotic behaviors, the number of stable and unstable photon spheres satisfies the relation $n_{\text{stable}} - n_{\text{unstable}} = -1$. In this study, we provide a new proof on this relation using a geometric analysis, which is carried out using intrinsic curvatures in the optical geometry of black hole spacetimes. Firstly, we demonstrate the existence of photon spheres near black holes assuming most general asymptotic behaviors (asymptotically flat black holes, asymptotically de-Sitter and anti-de-Sitter black holes). Subsequently, we prove that the stable and unstable photon spheres near black holes must be one-to-one alternatively separated from each other, such that each unstable photon sphere is sandwiched between two stable photon spheres (and each stable photon sphere is sandwiched between two unstable photon spheres). Our analysis is applicable to any spherically symmetric black hole spacetimes.
\end{abstract}


\maketitle

\section{Introduction \label{section1}}

The photon spheres / circular photon orbits are significantly important topics in black hole physics. On one hand, these photon spheres are closely connected with a number of astrophysical observations of black holes in galaxies, such as the gravitational lensing, black hole's accretion processes and black hole shadows. Particularly, since the observation of high-resolution images of supermassive black holes captured by Event Horizon Telescope (EHT) \cite{ETH2019a,ETH2022}, there has been large interests in studying the photon spheres and black hole shadows. Numerous studies have emerged to extensively investigate the photon spheres and shadow images for various types of black holes \cite{Johannsen2013,Hod2013,Tsukamoto2014,Atamurotov2015,Tsukamoto2018,Perlick2018,JiaJJ2018a,JiaJJ2018b,Shaikh2018,Mishra2019,Visinelli2019,Konoplya2019,WeiSW2019,Gralla2019,Joshi2020,LiuHS2019,ZhuQH2020,ZengXX2020,GuoM2020,Lima2021,GanQY2021a,GanQY2021b,GaoSJ2022,Adler2022,Perlick2022a,WangMZ2022,GuoGZ2023,Chen2023,Vagnozzi2023,Ghosh2023,Bargueno2023,Tsukamoto2024,Vertogradov2024,Murk2024}. On the other hand, photon spheres may also provide valuable insights into the physical properties and characteristics of black holes. They can reveal non-trivial aspects related to the spacetime singularities, event horizons, causal structures and asymptotic behaviors associated with black holes and ultracompact objects. For instance, a number of recent researches have pointed out that the number and distribution of photon spheres (or light rings) in black hole spacetimes, horizon-less spacetimes produced by ultracompact objects, and naked singularity spacetimes could exhibit entirely different features \cite{Cederbaum2016,Gibbons2016,Berry2020,Wielgus2021,Isomura2023,Cunha2017,Cunha2020}.

Conventionally, the photon spheres in the vicinity of black holes are obtained from the effective potential of photons moving in black hole spacetimes \cite{Hod2011,Carroll,Hartle,Straumann,Perlick2022a,Raffaelli2021,Destounis2023}. While this approach proves highly valuable for analytical and numerical investigations for a specific spacetime, it is advantageous to incorporate geometric and topological approaches, especially when seeking the universal properties associated with photon spheres. The geometric and topological techniques could provide us powerful tools on the studies of photon spheres, without having to solve effective potentials and particle orbits for any particular black hole spacetimes.

\begin{table*}
	\caption{Features of the conventional effective potential approach, the topological approach, and our geometric approach.}
	\label{table1}
	\vspace{2mm}
	\begin{ruledtabular}
		\begin{tabular}{lccccc}
			Approach & Our Geometric Approach & Topological Approach & Conventional Approach &
			\\
			\hline
			Geometry & Optical geometry       & Spacetime geometry   & Spacetime geometry &
			\\
			\hline
			Key quantities & Gaussian curvature $\mathcal{K}(r)$ & Normalized Vector Field $V$ & Effective Potential $V_{\text{eff}}(r)$ &
			\\
			& Geodesic curvature $\kappa_{g}(r)$ & Topological Charge $w$ &       
			\\
			\hline
			photon sphere  & $\kappa_{g}(r)=0$ & $V=0$ & $\frac{dV_{\text{eff}}(r)}{dr}=0$
			\\
			unstable photon sphere & $\kappa_{g}(r)=0$ and $\mathcal{K}(r)<0$ & $V=0$ and $w=-1$ & $\frac{dV_{\text{eff}}(r)}{dr}=0$ and $\frac{d^{2}V_{\text{eff}}(r)}{dr^{2}}<0$ &
			\\
			stable photon sphere & $\kappa_{g}(r)=0$ and $\mathcal{K}(r)>0$ & $V=0$ and $w=+1$ & $\frac{dV_{\text{eff}}(r)}{dr}=0$ and $\frac{d^{2}V_{\text{eff}}(r)}{dr^{2}}>0$ &
			\\
		\end{tabular}
	\end{ruledtabular}
\end{table*}

Under these circumstances, new approaches using topological invariant (such as topological charge, index of vector fields, and Brouwer's degree of mapping) and other mathematical concepts in the studying of photon spheres (or light rings) have already been stimulated \cite{Gibbons1993,Cunha2017,Cunha2018,Cunha2020,WeiSW2020,Virbhadra2001,Koga2019,Kobialko2022,SongY2023}. Notably, in 2017, P. V. P. Cunha \emph{et al.} introduced a topological approach to photon spheres / light rings, assigning a topological charge to each photon sphere (or light ring) in spacetime using the topological index of auxiliary vector fields \cite{Cunha2017}. Within this framework, they successfully demonstrated the existence of photon spheres in black hole spacetimes, and they also constrained the number of photon spheres (the number of stable and unstable photon spheres satisfies $n_{\text{stable}} - n_{\text{unstable}} = -1$ in black hole spacetimes)  \cite{Cunha2020}. Subsequently, this topological approach has been further developed to a wider range of gravitational systems, extending its utility to both photon spheres / light rings and timelike circular orbits \cite{WeiSW2020,WeiSW2023,WeiSW2023b,WeiSW2023c,YinJ2023,Lima2022,Tavlayan2023,Sadeghi2024,Cunha2024,Xavier2024,Afshar2024,Afshar2024b}. Additionally, S. -W. Wei \emph{et al.} pointed out that the topological charge can also be used to investigate the topology of black hole spacetimes as well as the black hole thermodynamic properties \cite{WeiSW2022,WeiSW2022b}. These new approaches could provide us new insights on black hole physics and spacetime geometry, potentially providing answers to some important questions about photon spheres, spacetime properties and gravitational fields.

Stimulated from these approaches and modern geometric techniques, we proposed a geometric approach to study photon spheres and black hole shadows for spherically symmetric black holes in recent works \cite{Qiao2022a,Qiao2022b}. In our geometric approach, the photon spheres / circular photon orbits are analyzed through the optical geometry of black hole spacetime, which can be constructed from a generalization of the Fermat's principle in curved spacetime \cite{Abramowicz1988,Gibbons2008,Gibbons2009,Werner2012,footnote-Fermat}. The photon orbits (traveling along null geodesics in spacetime geometry) retain geodesic when they are transformed into optical geometry. Two intrinsic curvatures in the optical geometry --- Gaussian curvature and geodesic curvature --- can completely determine the stable and unstable photon spheres. Notably, for any photon spheres, the geodesic curvature is precisely zero, namely $\kappa_{g}(r=r_{ph})=0$. The sign of Gaussian curvature identifies the stability of photon spheres. The negative Gaussian curvature signifies the corresponding photon spheres are unstable, while the positive Gaussian curvature indicates the photon spheres are stable. In addition, it has been proven that our geometric approach is completely equivalent to the conventional approach based on effective potential of test particles \cite{Qiao2022a,Qiao2022b}. Furthermore, this geometric approach can be applied to study the orbits of other particles. Particularly, P. V. P. Cunha \emph{et al.} extended our approach to study the timelike circular geodesics using the Jocabi geometry of black hole spacetimes \cite{Cunha2022}. The distinguishing features of the conventional approach (employing the effective potential of test particles moving in spacetimes), topological approach (utilizing the topological charge of photon spheres / topological index of given auxiliary vector fields) and our geometric approach (relying on the Gaussian curvature and geodesic curvature in the optical / Jacobi geometry of black hole spacetimes) are summarized in table \ref{table1}.

In the current study, we utilize a novel geometric analysis on the existence of photon spheres as well as the distributions of stable and unstable photon spheres in the vicinity of black holes. Because of the astrophysical observations of black hole shadow images, it is widely accepted that the existence of photon sphere must be a general feature in black hole spacetimes. By assuming the proper asymptotic behaviors of black hole spacetime, this conjecture has been proven from several different schemes \cite{JiaJJ2018a,JiaJJ2018b,Cederbaum2016,Cunha2020,GaoSJ2021,Ghosh2021}. In this work, we provide a new proof on the existence of photon spheres in asymptotically flat, asymptotically de-Sitter and anti de-Sitter black hole spacetimes, following our geometric approach proposed in references \cite{Qiao2022a,Qiao2022b}. Subsequently, we employ the Gauss-Bonnet theorem to show that the stable and unstable photon spheres near black holes must be one-to-one alternatively separated from each other, such that each unstable photon sphere is sandwiched between two stable photon spheres (and each stable photon sphere is sandwiched between two unstable photon spheres). This photon sphere distribution characteristic enables us to give a new demonstration of the important relation $n_{\text{stable}} - n_{\text{unstable}} = -1$ in black hole spacetimes from our geometric analysis. In the presented work, our analysis is confined to spherically symmetric black holes (with the spacetime metric to be $ds^{2}=g_{tt}dt^{2}+g_{rr}dr^{2}+g_{\theta\theta}d\theta^{2}+g_{\phi\phi}d\phi^{2}$).

This work is organized in the following way. Section \ref{section1} provides the backgrounds and motivations of our work. Section \ref{section2} offers a concise review of our geometric approach for analyzing photon spheres. Section \ref{section3} presents the proof on the existence of photon spheres in the vicinity of black holes (assuming the proper asymptotic behaviors of black holes spacetimes). Section \ref{section4} discusses in detail the distribution of stable and unstable photon spheres near black holes. In this section, we provide a new proof on the number of photon spheres $n_{\text{stable}} - n_{\text{unstable}} = -1$. The summary and perspectives are outlined in section \ref{section5}. In addition, some mathematical preliminaries of our work are described in the Appendices, including the optical geometry of black hole spacetime, two intrinsic curvatures in optical geometry (Gaussian curvature and geodesic curvature), the Gauss-Bonnet theorem in differential geometry, the connection between Gaussian curvature and the stability of photon spheres.

\section{Geometric Approach to Photon Spheres \label{section2}}

In this section, we briefly describe the geometric approach to photon spheres, which was proposed in our recent works \cite{Qiao2022a,Qiao2022b}. In this approach, a low-dimensional analog of the black hole spacetime --- the optical geometry --- is implemented. Intrinsic curvatures in optical geometry are capable to determine the locations of photon spheres and their stability in the vicinity of black holes. 
This approach is applicable to any spherically symmetric spacetime, and it could give completely equivalent results with the conventional geodesic approach based on effective potential of massless photons.

The optical geometry serves as a low-dimensional analog of the spacetime when describing the massless particle's orbits, providing powerful tools to study the motions of photons (or other massless particles traveling along null geodesics) in the gravitational field  \cite{Abramowicz1988,Gibbons2008,Gibbons2009,Werner2012}. The underlying physical interpretation of the optical geometry can be viewed as a generalization of the Fermat's principle in curved spacetime. In recent years, the optical geometry has attracted significant interests, and various aspects of black hole physics can be explored using the geometric and topological properties of optical geometry \cite{Abramowicz1988,Gibbons2008,Gibbons2009}. For instance, G. W. Gibbons and M. C. Werner developed an approach to calculate the gravitational deflection angle using the Gauss-Bonnet theorem in optical geometry \cite{Gibbons2008}. The Gibbons-Werner approach has been widely applied in a variety of gravitational systems and several modified versions of the Gibbons-Werner approach have also been developed \cite{Ishihara2016a,Ishihara2016b,Ono2017,Jusufi2018,Jusufi2018b,Crisnejo2018,Ono2019,Takizawa2020,LiZH2020,LiZH2020a,LiZH2020b,Huang2022,Huang2023,Takizawa2023,LiZH2024a,LiZH2024b}.

Mathematically, the optical geometry can be constructed in several equivalent ways. A straightforward method to construct the optical geometry is from the spacetime geometry through a continuous transformation with the null constraint $d\tau^{2}=-ds^{2}=0$ 
\begin{equation}
	\underbrace{ds^{2} = g_{\mu\nu}dx^{\mu}dx^{\nu}}_{\text{Spacetime Geometry}}
	\ \ \overset{d\tau^{2}=-ds^{2}=0}{\Longrightarrow} \ \ 
	\underbrace{dt^{2} = g^{\text{OP}}_{ij}dx^{i}dx^{j}}_{\text{Optical Geometry}}
	\label{optical geometry}
\end{equation}
Furthermore, for spherically symmetric black holes, the photon sphere can always be constricted in the equatorial plane without loss of generality. In this way, a 2-dimensional optical geometry can be constructed.
\begin{equation}
	\underbrace{dt^{2} = g^{\text{OP}}_{ij}dx^{i}dx^{j}}_{\text{Optical Geometry}}
	\ \ \overset{\theta=\pi/2}{\Longrightarrow} \ \ 
	\underbrace{dt^{2}=\tilde{g}^{\text{OP-2d}}_{ij}dx^{i}dx^{j}}_{\text{Optical Geometry (Two Dimensional)}}
	\label{optical geometry2}
\end{equation}
The detailed construction and explicit metric form of the optical geometry for spherically symmetric black hole spacetimes and rotational symmetric black spacetimes are presented in Appendix \ref{appendix1}. Notably, in the context of spherically symmetric black hole spacetimes, the corresponding optical geometry $dt^{2}=\tilde{g}^{\text{OP-2d}}_{ij}dx^{i}dx^{j}$ is a 2-dimensional Riemannian geometry \cite{Gibbons2008,Gibbons2009}.

The optical geometry of black hole spacetime has some desirable properties. The photon orbits, which follow null geodesic curves in spacetime geometry (characterized by a 4-dimensional Lorentz geometry), are converted to spatial geodesic curves when transforming into optical geometry. In other words, the null geodesic curve $\gamma=\gamma(\tau)$ in spacetime geometry $ds^{2}=g_{\mu\nu}dx^{\mu}dx^{\nu}$ retains geodesic in the optical geometry $dt^{2}=g^{\text{OP}}_{ij}dx^{i}dx^{j}$. Additionally, the optical geometry possesses simpler intrinsic geometric and topological properties than the 4-dimensional spacetime geometry (because the dimension is reduced). In this low-dimensional geometry, a number of elegant theorems in classical surface theory, modern differential geometry and topology could offer powerful tools to study the particle motions in the vicinity of black holes. 

In our geometric approach developed in references \cite{Qiao2022a,Qiao2022b}, the analysis of photon spheres / circular photon orbits and their stability is implemented in the 2-dimensional equatorial plane of optical geometry (obtained from equation (\ref{optical geometry2})). The most important intrinsic curvatures in this 2-dimensional optical geometry are Gaussian curvature and geodesic curvature, which are explained in Appendix \ref{appendix2}. The Gaussian curvature is the curvature of a 2-dimensional surface $S$, quantifying how much this surface deviates from being intrinsically flat. The geodesic curvature, on the other hand, describes the curvature of a continuous curve $\gamma(s)$, which measures how far this curve departs from being a geodesic curve on this surface \cite{Carmo1976,Berger,Berger2}. Using the Gaussian curvature and geodesic curvature in optical geometry, the locations of photon spheres near black holes and their stability can be completely determined. Firstly, the photon sphere $r=r_{ph}$ near the black hole, which is a geodesic curve in spacetime geometry, becomes a spatial geodesic curve in the 2-dimensional optical geometry, and its geodesic curvature vanishes naturally \cite{Qiao2022a,Qiao2022b}
\begin{equation} 
	\kappa_{g}(r=r_{ph})  =  0 \ .
\end{equation}
The solution of this equation gives the exact position of photon spheres in the vicinity of black holes. Further, it is also demonstrated that the above geodesic curvature criterion in our geometric approach is equivalent to the effective potential criterion within the conventional approach \cite{Qiao2022a,Qiao2022b}.
\begin{equation} 
	\kappa_{g}(r=r_{\text{ph}})  =  0  
	\ \ \Leftrightarrow \ \
	\frac{dV_{\text{eff}}(r)}{dr} \bigg|_{r=r_{\text{ph}}} = 0 
\end{equation}
Secondly, the stability of photon spheres near black holes can be constrained by the topological properties of optical geometry, especially the existence of conjugate points. For stable and unstable photon spheres, the behavior of photon orbits under a perturbation from photon spheres are completely different. When photons are perturbed at a given point on the unstable photon sphere, they would either escape to infinity or fall into black holes. Conversely, when photons are perturbed from a stable photon sphere, apart from going to infinity or falling into black holes, they could also travel along another bound photon orbit near the original photon sphere. Mathematically, these distinct characteristics of stable and unstable photon sphere are reflected by conjugate points in the manifold. There are conjugate points in the stable photon sphere, while no conjugate points could exist in the unstable photon sphere. Thus, the existence or absence of conjugate points offer us a novel means to distinguish the stable and unstable photon spheres. A classical theorem in differential geometry and topology --- Cartan-Hadamard theorem --- strongly constrain the Gaussian curvature and the existence of conjugate points. By applying the Cartan-Hadamard theorem in the equatorial plane of optical geometry, we derive the following criterion to determine the stability of photon spheres / circular photon orbits \cite{Qiao2022a,Qiao2022b}.
\begin{eqnarray}
	\mathcal{K}(r) < 0 & \Rightarrow & \text{The photon shere $r=r_{ph}$ is unstable} \nonumber 
	\\
	\mathcal{K}(r) > 0 & \Rightarrow & \text{The photon sphere $r=r_{ph}$ is stable} \nonumber
\end{eqnarray}
The detailed interpretation on the relationship between Gaussian curvature, stability of photon spheres and the Cartan-Hadamard theorem is presented in Appendix \ref{appendix4}. This conclusion, derived purely from the geometric and topological properties of optical geometry, is independent of the specific metric form of black hole spacetimes (only constraining the sign of Gaussian curvature). Consequently, the analysis can be universally applicable to any spherically spherical black hole. Furthermore, it has been demonstrated that the Gaussian curvature criterion for stable and unstable photon spheres is equivalent to the effective potential criterion within the conventional approach \cite{Qiao2022a,Qiao2022b}
\begin{subequations} 
\begin{eqnarray}
		\frac{d^{2}V_{\text{eff}}(r)}{dr^{2}} \bigg|_{r=r_{\text{unstable}}} < 0 
		\ \ & \Leftrightarrow & \ \ 
		\mathcal{K}(r=r_{\text{unstable}}) < 0 \ \ \ \ \ \ \ \  
		\\
		\frac{d^{2}V_{\text{eff}}(r)}{dr^{2}} \bigg|_{r=r_{\text{stable}}} > 0 
		\ \ & \Leftrightarrow & \ \ 
		\mathcal{K}(r=r_{\text{stable}}) > 0
\end{eqnarray}
\end{subequations}
The distinguishing features of our geometric approach, the conventional effective potential approach, the topological approach and their equivalence have been summarized in table \ref{table1}.

\begin{widetext}
In the following, we present the detailed expression of the geodesic curvature and Gaussian curvature in the optical geometry of spherically symmetric black holes. The spacetime metric for general static spherically symmetric black holes can be expressed as
\begin{equation}
	ds^{2}=g_{tt}(r)dt^{2}+g_{rr}(r)dr^{2}+g_{\theta\theta}(r)d\theta^{2}+g_{\phi\phi}(r,\theta)d\phi^{2} .
	\label{spacetime metric}
\end{equation}
The corresponding optical geometry restricted in the equatorial plane $\theta=\pi/2$ is given by 
\begin{eqnarray}
	dt^{2} & = & \tilde{g}^{\text{OP-2d}}_{ij}dx^{i}dx^{j} 
	= \tilde{g}^{\text{OP-2d}}_{rr}dr^{2} + \tilde{g}^{\text{OP-2d}}_{\phi\phi}d\phi^{2} 
	= - \frac{g_{rr}(r)}{g_{tt}(r)} \cdot dr^{2} - \frac{\overline{g}_{\phi\phi}(r)}{g_{tt}(r)} \cdot d\phi^{2} ,
\end{eqnarray}
where $\overline{g}_{\phi\phi}(r)=g_{\phi\phi}(r,\theta=\pi/2)$. The geodesic curvature of a circle with constant radius $r$ (such as the photon spheres) in the 2-dimensional optical geometry can be calculated via
\begin{eqnarray}
	\kappa_{g}(\gamma = C_{r}) = \kappa_{g}(r) 
	& = & \frac{1}{2\sqrt{\tilde{g}^{\text{OP-2d}}_{rr}}} 
	      \frac{\partial \big[ \text{log}(\tilde{g}^{\text{OP-2d}}_{\phi\phi})\big]}{\partial r} 
	= \frac{1}{2\sqrt{\tilde{g}^{\text{OP-2d}}}} \cdot
	  \frac{1}{\sqrt{\tilde{g}^{\text{OP-2d}}_{\phi\phi}}}
	  \frac{\partial \tilde{g}^{\text{OP-2d}}_{\phi\phi}}{\partial r} 
	\nonumber
	\\
	& = &
	\frac{g_{tt}(r)}{2\overline{g}_{\phi\phi}(r)} \cdot \sqrt{-\frac{g_{tt}(r)}{g_{rr}(r)}} 
	\cdot \frac{\partial}{\partial r} 
	\bigg[ \frac{\overline{g}_{\phi\phi}(r)}{g_{tt}(r)} \bigg] ,
	\label{geodesic cuurvature expression}
\end{eqnarray}
and the Gaussian curvature in the 2-dimensional optical geometry can be calculated through 
\begin{eqnarray}
	\mathcal{K} & = & -\frac{1}{\sqrt{\tilde{g}^{\text{OP-2d}}}} 
	\bigg[
	\frac{\partial}{\partial \phi} \bigg( \frac{1}{\sqrt{\tilde{g}^{\text{OP-2d}}_{\phi\phi}}} \frac{\partial\sqrt{\tilde{g}^{\text{OP-2d}}_{rr}}}{\partial \phi}  \bigg)
	+ \frac{\partial}{\partial r} \bigg( \frac{1}{\sqrt{\tilde{g}^{\text{OP-2d}}_{rr}}} \frac{\partial\sqrt{\tilde{g}^{\text{OP-2d}}_{\phi\phi}}}{\partial r}  \bigg)
	\bigg]  \nonumber
	\\
	& = & \frac{g_{tt}(r)}{\sqrt{g_{rr}(r)\cdot \overline{g}_{\phi\phi}(r)}}
	      \cdot \frac{\partial}{\partial r}
	      \bigg[
	        \frac{g_{tt}(r)}{2\sqrt{g_{rr}(r)\cdot \overline{g}_{\phi\phi}(r)}}
	        \cdot \frac{\partial}{\partial r} \bigg( \frac{\overline{g}_{\phi\phi}(r)}{g_{tt}(r)} \bigg)
	      \bigg] , \label{Gauss cuurvature expression}
\end{eqnarray}
with $\tilde{g}^{\text{OP-2d}}=\text{det} \big( \tilde{g}^{\text{OP-2d}}_{ij} \big)$ to be the determinant of the 2-dimensional optical geometry metric.
\end{widetext}


\section{The Existence of Photon Spheres near Black Holes \label{section3}}

In this section, we prove the existence of photon spheres in the vicinity of spherically symmetric black holes. The proof is based on our geometric approach and the asymptotic behaviors of black holes, irrespective of any specific spacetime metric functions $g_{tt}(r)$ and $g_{rr}(r)$. In this study, we give the detailed proof for black holes with most commonly asymptotic behaviors --- the asymptotically flat, asymptotically de-Sitter and asymptotically anti de-Sitter black holes. However, as will be shown through the subsequent analysis, a similar analysis can be applied to a wide range class of black holes with other asymptotic behaviors.

\begin{figure}
	\includegraphics[width=0.5\textwidth]{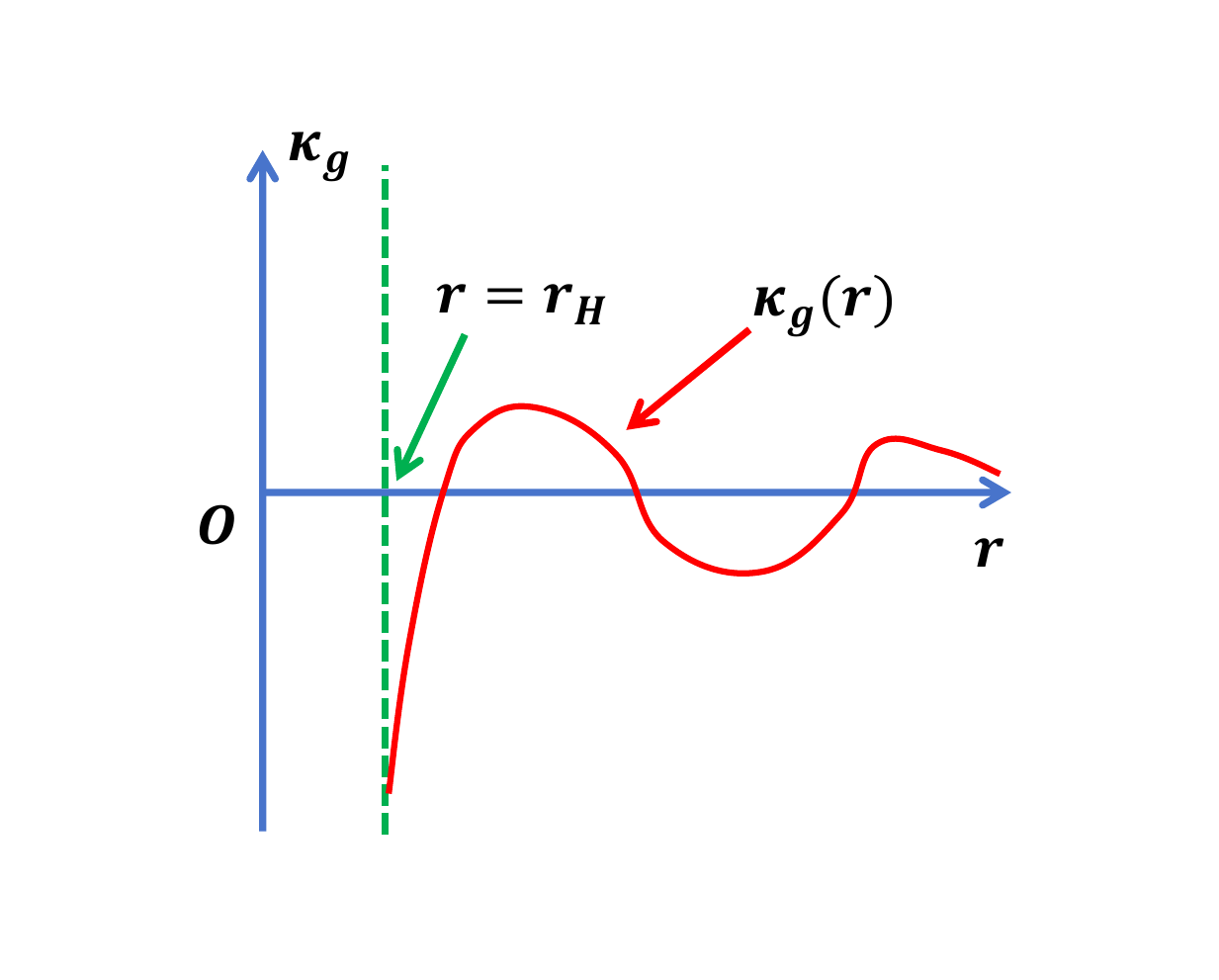}
	\caption{The variation of geodesic curvature $\kappa_{g}(r)$ in 2-dimensional optical geometry with respect to radial coordinate $r$. The geodesic curvature in the near horizon limit is related to the surface gravity of black holes, via $\lim_{r \to r_{H}}\kappa_{g}(r)=-\kappa_{\text{surface}} < 0$. If the geodesic curvature in the infinite distance limit satisfies $\lim_{r \to \infty}\kappa_{g}(r) > 0$, the equation $\kappa_{g}(r)=0$ must have at least one solution.}
	\label{figure1}
\end{figure} 

In our geometric analysis, the criterion for the presence of photon spheres is the vanishing of geodesic curvature, namely $\kappa_{g}(r=r_{ph})=0$. Therefore, the existence of photon spheres is equivalent to that the equation $\kappa_{g}(r)=0$ admit at least one solution \cite{footnote notation}. To prove the existence of photon spheres near black holes, the crucial point is analyzing the behavior of geodesic curvature for a circular curve in the 2-dimensional optical geometry, especially in the near horizon region and the infinite distance region. Notably, for the spacetimes generated by black holes, a recent study have shown that the geodesic curvature $\kappa_{g}(r)$ in the optical geometry in near horizon limit is related to the surface gravity of black holes via $\lim_{r \to r_{H}}\kappa_{g}(r)=-\kappa_{\text{surface}} < 0$ \cite{Cunha2022}, which can be viewed as a general feature of optical geometry. Under these circumstances, if the geodesic curvature satisfies $\lim_{r \to \infty} \kappa_{g}(r)>0$ in the infinite distance limit, as illustrated in figure \ref{figure1}, then the equation $\kappa_{g}(r)=0$ must have at least one solution, thereby proving the existence of photon spheres in black hole spacetimes. In the following, we derive the infinite distance limit $\lim_{r \to \infty} \kappa_{g}(r)>0$ for the asymptotically flat black holes, asymptotically de-Sitter and asymptotically anti de-Sitter black holes, respectively.

\textbf{Asymptotically Flat Black Holes:}
The asymptotically flat black holes has the following analytical metric expansions in the $r \to \infty$ limit
\begin{subequations}
\begin{align}
	\lim_{r \to \infty} g_{tt}(r) & = - \bigg[ 1 + \frac{a_{i}}{r^{i}} + O\bigg( \frac{1}{r^{i+1}} \bigg) \bigg] & (i \ge 1)
	\\
	\lim_{r \to \infty} g_{rr}(r) & = \ \ 1 - \frac{b_{j}}{r^{j}} + O\bigg( \frac{1}{r^{j+1}} \bigg) & (j \ge 1)
\end{align}
\end{subequations}
with $a_{i}$ and $b_{j}$ to be the characteristic expansion coefficients. Furthermore, for any spherically symmetric black holes, the metric components $g_{\theta\theta}(r)=r^{2}$ and $g_{\phi\phi}(r)=r^{2}\sin^{2}\theta$ can be achieved through a coordinate transformation. Using these metric expansions, we eventually obtain the geodesic curvature for a circular curve in the infinite distance limit
\begin{eqnarray}
	\lim_{r\to\infty}\kappa_{g}(r)
	& = &
	\lim_{r\to\infty} 
	\bigg[
	\frac{g_{tt}(r)}{2\overline{g}_{\phi\phi}(r)} \cdot \sqrt{-\frac{g_{tt}(r)}{g_{rr}(r)}} 
	\cdot \frac{\partial}{\partial r} \bigg( \frac{\overline{g}_{\phi\phi}(r)}{g_{tt}(r)} \bigg)
	\bigg] \nonumber
	\\
	& = & \lim_{r\to\infty}
	\frac{\frac{1}{r}+\frac{a_{i}}{r^{i+1}}+\frac{ia_{i}}{2r^{i+1}}+O\big(\frac{1}{r^{i+2}}\big)}{\sqrt{\bigg[1+\frac{a_{i}}{r^{i}}+O\big(\frac{1}{r^{i+1}}\big)\bigg]\bigg[1-\frac{b_{j}}{r^{j}}+O\big(\frac{1}{r^{j+1}}\big)\bigg]}} \nonumber
	\\
	& = & \lim_{r\to\infty} \frac{1}{r} = 0^{+} 
\end{eqnarray}
The notation $\lim_{r\to\infty}\kappa_{g}(r)=0^{+}$ in this expression indicates that the geodesic curvature satisfies $\kappa_{g}(r) \to 0$ and $\kappa_{g}(r)>0$ in the $r \to \infty$ limit. This is exactly the case illustrated in figure \ref{figure1}, suggesting that the equation $\kappa_{g}(r)=0$ has at least one solution. Consequently, the existence of photon spheres is demonstrated for asymptotically flat black hole spacetimes.

\textbf{Asymptotically de Sitter and Asymptotically Anti de-Sitter Black Holes:}
In the infinite distance limit, the asymptotically de Sitter / asymptotically anti de-Sitter black holes have the asymptotic analytical metric expansions
\begin{subequations}
\begin{align}
	\lim_{r \to \infty} g_{tt}(r) & = - \bigg[ 1 - \frac{\Lambda r^{2}}{3} + \frac{a_{i}}{r^{i}} + O\bigg( \frac{1}{r^{i+1}} \bigg) \bigg] & (i \ge 1) 
	\\
	\lim_{r \to \infty} g_{rr}(r) & = \ \ \frac{1}{1-\frac{\Lambda r^{2}}{3}+\frac{b_{j}}{r^{j}}+O\big(\frac{1}{r^{j+1}}\big)} & (j \ge 1)
\end{align}
\end{subequations}
with $a_{i}$ and $b_{j}$ to be the characteristic expansion coefficients, and $\Lambda$ is the cosmological constant. The asymptotically de Sitter black holes admit a positive cosmological constant $\Lambda>0$, while the asymptotically anti de Sitter black holes have a negative cosmological constant $\Lambda<0$. Similarly, since we are focused on the spherically symmetric black holes, the metric components $g_{\theta\theta}(r)=r^{2}$ and $g_{\phi\phi}(r)=r^{2}\sin^{2}\theta$ can always be achieved through a coordinate transformation. Using the above metric expansions, we obtain the geodesic curvature in the infinite distance limit
\begin{eqnarray}
	\lim_{r\to\infty}\kappa_{g}(r)
	& = &
	\lim_{r\to\infty} 
	\bigg[
	\frac{g_{tt}(r)}{2\overline{g}_{\phi\phi}(r)} \cdot \sqrt{-\frac{g_{tt}(r)}{g_{rr}(r)}} 
	\cdot \frac{\partial}{\partial r} \bigg( \frac{\overline{g}_{\phi\phi}(r)}{g_{tt}(r)} \bigg)
	\bigg] \nonumber
	\\
    & = & \lim_{r\to\infty}
          \frac{ \frac{1}{r} + \frac{a_{i}}{r^{i+1}} + \frac{ia_{i}}{2r^{i+1}} + O\big(\frac{1}{r^{i+2}}\big) }
          { \sqrt{\frac{1-\frac{\Lambda r^{2}}{3}+\frac{a_{i}}{r^{i}}+O\big(\frac{1}{r^{i+1}}\big)}{1-\frac{\Lambda r^{2}}{3}+\frac{b_{j}}{r^{j}}+O\big(\frac{1}{r^{j+1}}\big)} } } \nonumber
    \\
    & = & \lim_{r\to\infty}
          \frac{ \frac{1}{r} + \frac{a_{i}}{r^{i+1}} + \frac{ia_{i}}{2r^{i+1}} + O\big(\frac{1}{r^{i+2}}\big) }
          { \sqrt{ 1 + \frac{\frac{a_{i}}{r^{i}}-\frac{b_{j}}{r^{j}}+O\big(\frac{1}{r^{i+1}}, \frac{1}{r^{j+1}}\big)}{1-\frac{\Lambda r^{2}}{3}+\frac{b_{j}}{r^{j}}+O\big(\frac{1}{r^{j+1}}\big)} } } \nonumber
    \\
    & = & \lim_{r\to\infty} \frac{1}{r} = 0^{+} 
\end{eqnarray}
The notation $\lim_{r\to\infty}\kappa_{g}(r)=0^{+}$ indicates that the geodesic curvature satisfies $\kappa_{g}(r) \to 0$ and $\kappa_{g}(r)>0$ in the $r \to \infty$ limit, which is precisely the same with those for asymptotically flat black holes. This successfully reproduces the case illustrated in figure \ref{figure1}, thus the equation $\kappa_{g}(r)=0$ must have at least one solution. The existence of photon spheres is demonstrated for both asymptotically de Sitter and asymptotically anti de-Sitter black holes.

In conclusion, for asymptotically flat, asymptotically de-Sitter and asymptotically anti de-Sitter black holes, the geodesic curvatures all satisfy $\lim_{r \to r_{H}}\kappa_{g}(r) < 0$ in the near horizon region and $\lim_{r\to\infty}\kappa_{g}(r) = 0^{+}$ in the infinite distance region, which are all consistent with the behavior shown in figure \ref{figure1}. The indicates that the equation $\kappa_{g}(r)=0$ must have at least one solution, explicitly proving the existence of photon spheres near such black holes. Furthermore, by assuming the continuity of geodesic curvature $\kappa_{g}(r)$ \cite{footnote continuous}, it can also be easily derived that the total number of photon spheres in the vicinity of black holes must be odd, namely $n_{\text{stable}} + n_{\text{unstable}} = 2k+1$ (which is extremely helpful for providing a new proof of the relation $n_{\text{stable}} - n_{\text{unstable}} = -1$ in the next section). In the present study, we haven't presented the detailed calculation of geodesic curvature in the infinite distance limit for black holes with other asymptotic behaviors. However, the similar analysis can be easily extended to black holes with other asymptotic behaviors.

\section{Distribution of Photon Spheres \label{section4}}

\begin{figure}
	\includegraphics[width=0.475\textwidth]{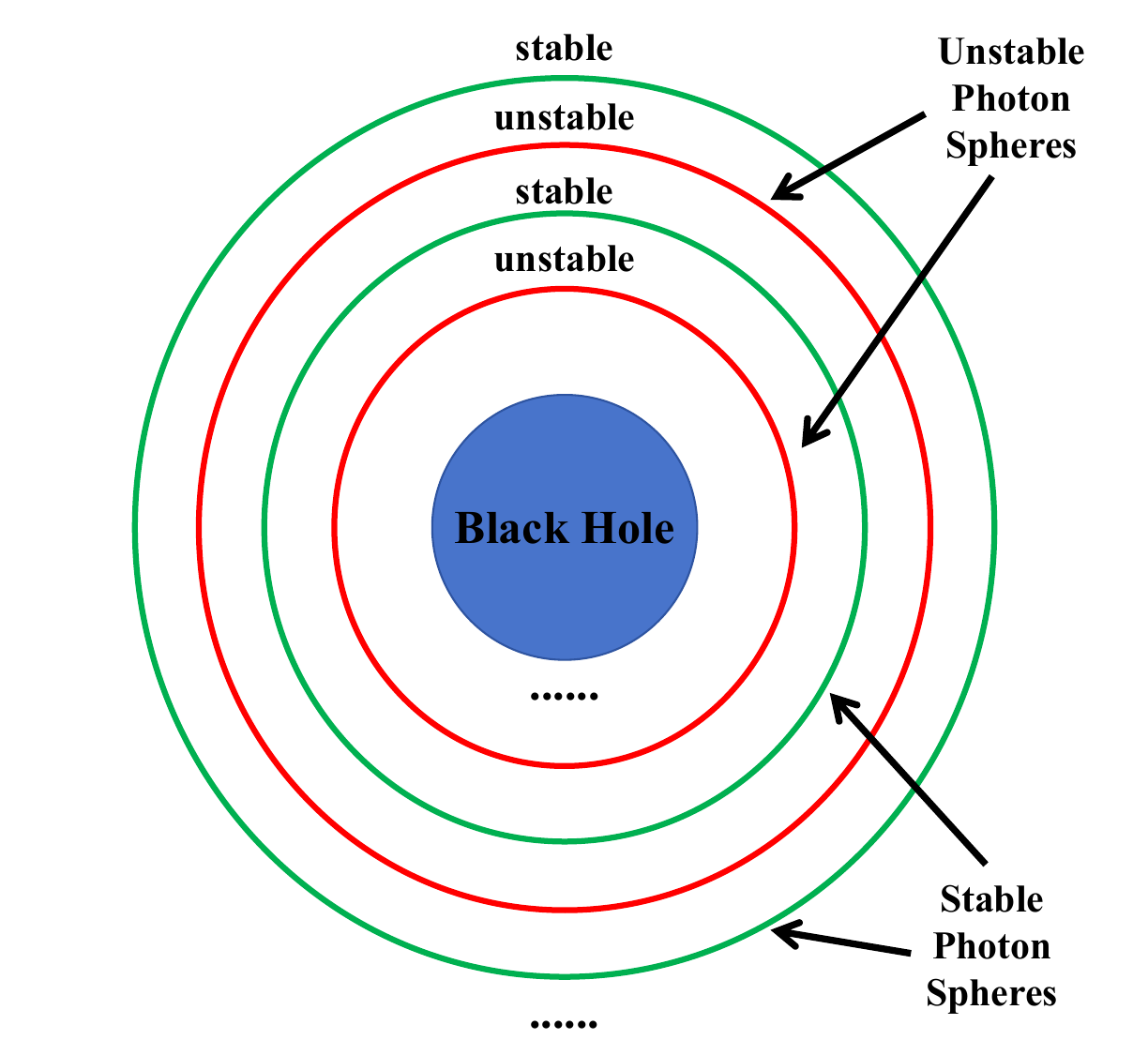}
	\caption{The stable and unstable photon spheres in black hole spacetimes are one-to-one alternatively separated from each other.}
	\label{figure2}
\end{figure}

In this section, we present a discussion on the distribution of photon spheres near spherically symmetric black holes. Based on our geometric analysis, the following property of photon sphere distribution can be demonstrated: 
\begin{quote}
    The stable and unstable photon spheres near black holes are one-to-one alternatively separated from each other, such that each unstable photon sphere is sandwiched between two stable photon spheres (and visa versa, each stable photon sphere is sandwiched between two unstable photon spheres), as illustrated in figure \ref{figure2}. 
\end{quote}
Furthermore, this distribution characteristic could lead to a new proof of a theorem on photon spheres proposed by Cunha \emph{et al.} recently \cite{Cunha2020}, which suggested the numbers of stable photon spheres and unstable photon spheres satisfy $n_{\text{stable}} - n_{\text{unstable}} = -1$. The $w = n_{\text{stable}} - n_{\text{unstable}} = -1$ can be viewed as topological invariant / topological charge for black hole spacetimes.

\begin{figure}
	\includegraphics[width=0.5\textwidth]{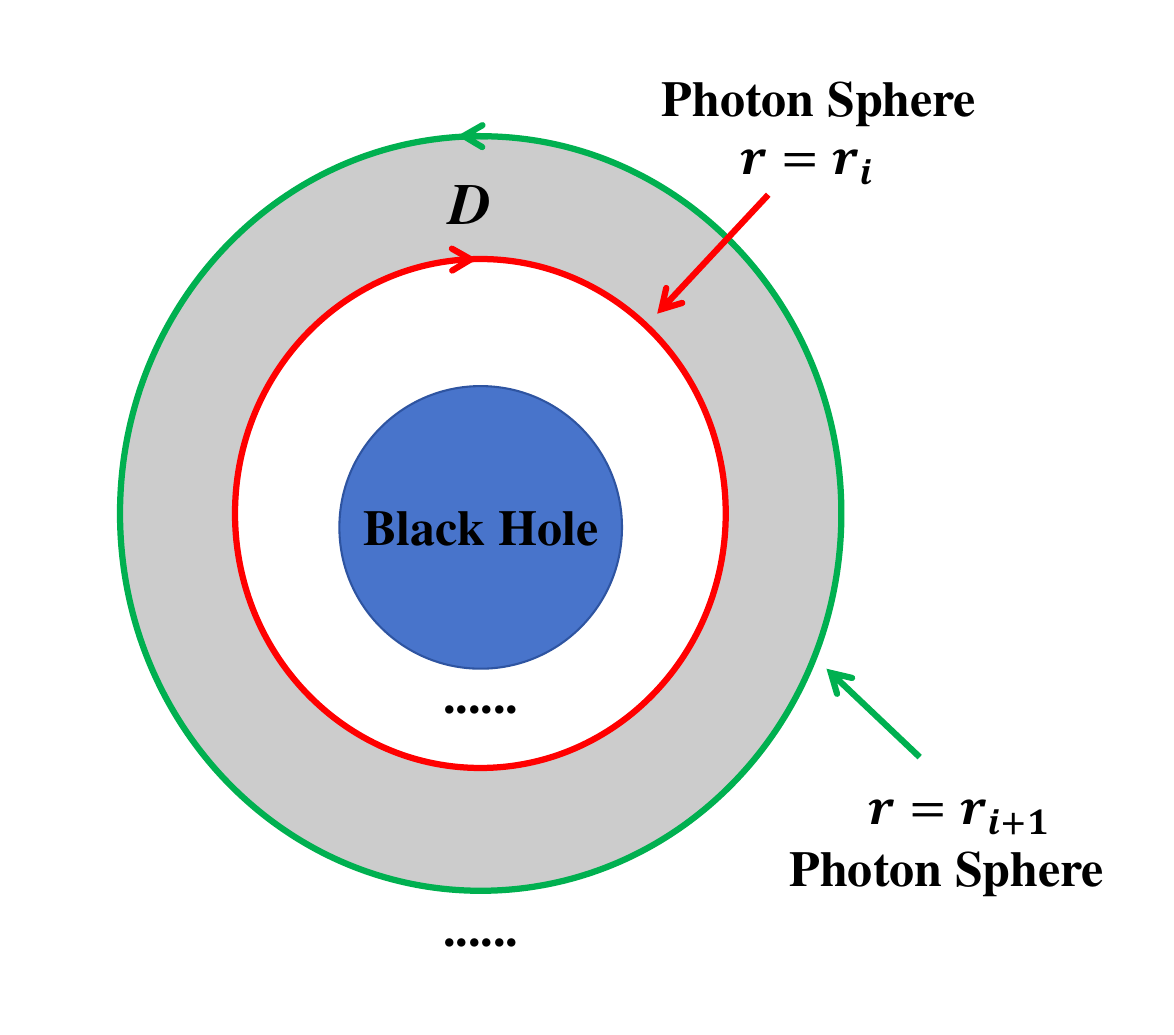}
	\caption{The choice of region $D$ in 2-dimensional optical geometry in the Gauss-Bonnet theorem to demonstrate that stable and unstable photon spheres in the vicinity of black hole are one-to-one alternatively separated from each other.}
	\label{figure3}
\end{figure}

In this study, instead of defining a topological charge $w$ relevant to photon spheres, our analysis is carried out from the Gauss-Bonnet theorem applied to optical geometry. Firstly, we select a region $D$ in the equatorial plane of optical geometry. The corresponding region $D$, which is shown in figure \ref{figure3}, is an annular zone bounded by two adjacent photon spheres without any other boundaries. It is a complexly connected region, such that the Euler characteristic number of this region is $\chi(D)=0$, which can be directly obtained from the triangulation of region $D$. The application of Gauss-Bonnet theorem within this region yields
\begin{equation}
	\int_{D}\mathcal{K}\cdot dS = 2 \pi \chi(D) = 0 \ ,
\end{equation}
which means the ``averaged'' Gaussian curvature between two adjacent photon spheres is zero.

In particular, inspired by recent works, a reduction on the surface integral of Gaussian curvature in the 2-dimensional optical geometry can be carried out \cite{Huang2022,Huang2023}. The Gauss-Bonnet theorem for the annular region $D$ eventually leads to
\begin{eqnarray}
	\int_{D}\mathcal{K}\cdot dS 
	& = & 
	\int_{0}^{2\pi}d\phi \int_{r_{i}}^{r_{i+1}} \mathcal{K}(r) \sqrt{\tilde{g}^{\text{OP-2d}}(r)} dr \nonumber
	\\
	& = & 2 \pi \big[ H(r_{i+1})-H(r_{i}) \big] = 0 \ .
	\label{Gauss-Bonnet reduce}
\end{eqnarray}
A simple derivation of the reduced formula (\ref{Gauss-Bonnet reduce}) is presented in Appendix \ref{appendix3}. Here, $H(r)$ is a function defined in reference \cite{Huang2022}
\begin{equation}
	H(r) = - \frac{1}{2\sqrt{\tilde{g}^{\text{OP-2d}}}}
	         \cdot \frac{\partial \tilde{g}^{\text{OP-2d}}_{\phi\phi}}{\partial r}
	     = - \sqrt{\tilde{g}^{\text{OP-2d}}_{\phi\phi}} \cdot \kappa_{g}(r) \ ,
	\label{function H}
\end{equation} 
with $\tilde{g}^{\text{OP-2d}}$ to be the determinant of 2-dimensional optical geometry metric. Notably, the last equality in (\ref{function H}) comes from the explicit expression of geodesic curvature for circles given in equation (\ref{geodesic cuurvature expression}). Particularly, both the inner and outer boundaries of region $D$ are photon spheres with vanishing geodesic curvature, so it is necessary to have $\kappa_{g}(r_{i+1}) = \kappa_{g}(r_{i}) = 0 \Rightarrow H(r_{i+1}) = H(r_{i}) = 0$ required for the selected annular region $D$.

\begin{figure*}
	\includegraphics[width=0.575\textwidth]{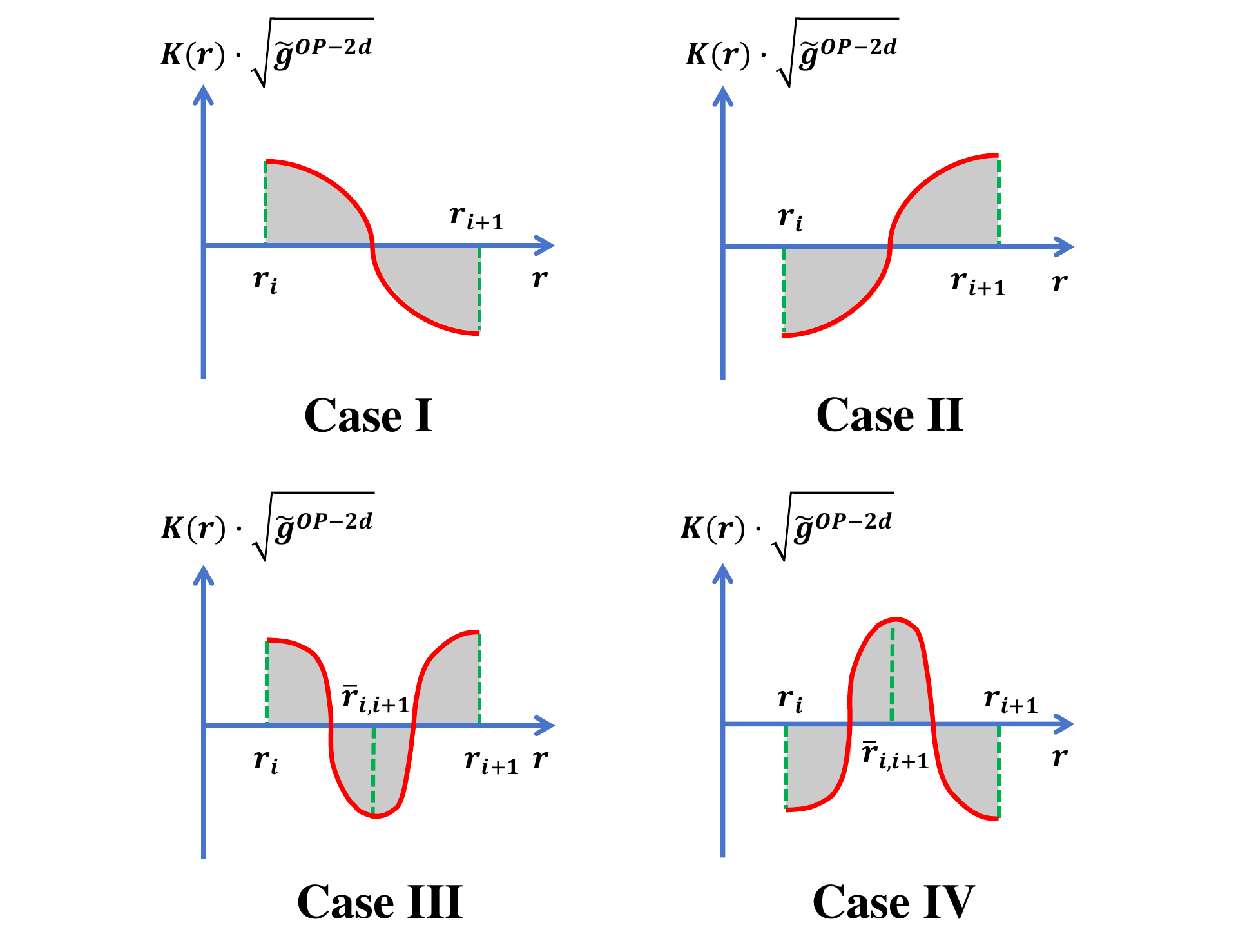}
	\caption{This figure summarizes the four different cases in which the surface integral of Gaussian curvature vanishes in region $D$ (namely $\int_{D}\mathcal{K}\cdot dS = 0$). \textbf{Case I:} the inner boundary photon sphere is stable and the outer boundary photon sphere is unstable, with $\mathcal{K}(r_{i}) > 0$ and $\mathcal{K}(r_{i+1}) < 0$. \textbf{Case II:} the inner boundary photon sphere is unstable and the outer boundary photon sphere is stable, with $\mathcal{K}(r_{i}) < 0$ and $\mathcal{K}(r_{i+1}) > 0$. \textbf{Case III:} both the inner and outer boundary photon spheres are stable, with $\mathcal{K}(r_{i}) > 0$ and $\mathcal{K}(r_{i+1}) > 0$. \textbf{Case IV:} both the inner and outer boundary photon spheres are unstable, with $\mathcal{K}(r_{i}) < 0$ and $\mathcal{K}(r_{i+1}) < 0$. In the last two cases, the inner and outer boundary photon spheres at $r=r_{i}$ and $r_{i+1}$ cannot be adjacent, and an additional photon sphere must exist between $r_{i}$ and $r_{i+1}$. The factor $\sqrt{\tilde{g}^{\text{OP-2d}}}$ in surface integral is always positive, so the sign of Gaussian curvature can be easily observed in this figure. This figure corresponds to the situations where Gaussian curvature is continuous in region $D$.}
	\label{figure4}
\end{figure*}

\begin{figure*}
	\includegraphics[width=0.575\textwidth]{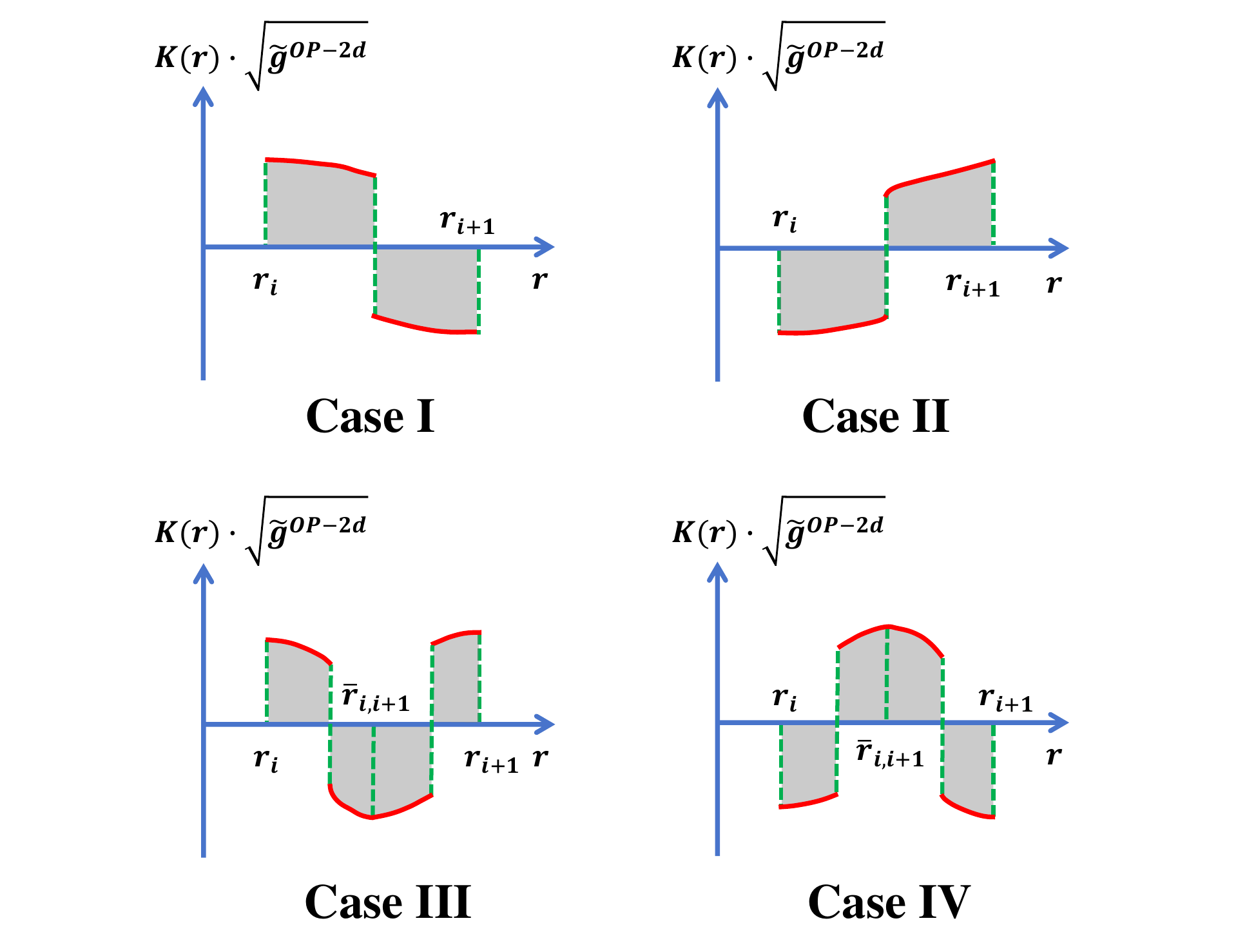}
	\caption{This figure summarizes the four different cases in which the surface integral of Gaussian curvature vanishes in region $D$ (namely $\int_{D}\mathcal{K}\cdot dS = 0$). \textbf{Case I:} the inner boundary photon sphere is stable and the outer boundary photon sphere is unstable, with $\mathcal{K}(r_{i}) > 0$ and $\mathcal{K}(r_{i+1}) < 0$. \textbf{Case II:} the inner boundary photon sphere is unstable and the outer boundary photon sphere is stable, with $\mathcal{K}(r_{i}) < 0$ and $\mathcal{K}(r_{i+1}) > 0$. \textbf{Case III:} both the inner and outer boundary photon spheres are stable, with $\mathcal{K}(r_{i}) > 0$ and $\mathcal{K}(r_{i+1}) > 0$. \textbf{Case IV:} both the inner and outer boundary photon spheres are unstable, with $\mathcal{K}(r_{i}) < 0$ and $\mathcal{K}(r_{i+1}) < 0$. In the last two cases, the inner and outer boundary photon spheres at $r_{i}$ and $r_{i+1}$ cannot be adjacent, and an additional photon sphere must exist between $r_{i}$ and $r_{i+1}$. The factor $\sqrt{\tilde{g}^{\text{OP-2d}}}$ in surface integral is always positive, so the sign of Gaussian curvature can be easily observed in this figure. This figure corresponds to the situations where Gaussian curvature admits finite number of discontinuous points with respect to radial coordinate $r$.}
	\label{figure5}
\end{figure*}

There are four different cases that could lead to the surface integral of Gaussian curvature vanishes in region $D$. These cases have been summarized in figure \ref{figure4}. The vanishing of surface integral $\int_{D}\mathcal{K}\cdot dS = 2 \pi \int_{r=r_{i}}^{r=r_{i+1}} \mathcal{K}(r) \sqrt{\tilde{g}^{\text{OP-2d}}} dr = 0$ means the shadow areas depicted in figure \ref{figure4} in the integration are zero. 
In the first two cases, the inner and outer photon spheres are of different types (one is a stable photon sphere, and the other is an unstable photon sphere), resulting in on-to-one alternating distributions of stable and unstable photon spheres, as depicted in figure \ref{figure2}. However, in the last two cases, the inner and outer photon spheres are of the same types, which violates the one-to-one alternatively separated distribution of stable and unstable photon spheres. Fortunately, a further analysis suggest that there must exist an additional photon sphere between the inner boundary photon sphere $r=r_{i}$ and outer boundary photon sphere $r=r_{i+1}$. Taking the case III as an example to show this conclusion. If the inner and outer photon spheres are both stable photon spheres with  $\mathcal{K}(r=r_{i}) > 0$ and $\mathcal{K}(r=r_{i+1}) > 0$, we can always find a position $\bar{r}_{i,i+1}$ such that the region $D$ is divided into two parts, demanding that the integration of Gaussian curvature $\int \mathcal{K}(r) \sqrt{\tilde{g}^{\text{OP-2d}}(r)} dr$ over intervals $[r_{i}, \bar{r}_{i,i+1}]$ and $[\bar{r}_{i,i+1}, r_{i+1}]$ both equal to zero (and the shadow areas in intervals $[r_{i}, \bar{r}_{i,i+1}]$ and $[\bar{r}_{i,i+1}, r_{i+1}]$ in the lower-left panel of figure \ref{figure4} are both zero). From the simplified integration formulas in (\ref{Gauss-Bonnet reduce}), it is evident that this position satisfies $H(\bar{r}_{i,i+1}) = H(r_{i}) = H(r_{i+1})$ and $\kappa_{g}(\bar{r}_{i,i+1}) = \kappa_{g}(r_{i+1}) = \kappa_{g}(r_{i}) = 0$, indicating $r=\bar{r}_{i,i+1}$ must be the location of a new photon sphere. Furthermore, the negative Gaussian curvature in lower-left panel of figure \ref{figure4} indicates that this new photon sphere at $r=\bar{r}_{i,i+1}$ is an unstable photon sphere, which is opposite to the inner and outer photon spheres at $r=r_{i}$, $r=r_{i+1}$ (both of them are stable photon spheres with positive Gaussian curvature). A similar analysis can be carried out for case IV, where an additional stable photon sphere must exist between the inner and outer unstable photon spheres at $r=r_{i}$ and $r=r_{i+1}$. This analysis remains valid even when there are finite discontinuous points in the Gaussian curvature with respect to radial coordinate $r$ (as presented in figure \ref{figure5}) \cite{footnote Gauss-Bonnet discontinuity}. In conclusions, among all four cases, we have proven that two stable photon spheres (or two unstable photon spheres) cannot be adjacent. The stable and unstable photon spheres in the vicinity of spherically symmetric black holes must be one-to-one alternatively separated from each other. Each stable photon sphere is sandwiched between two nearby unstable photon spheres, and each unstable photon sphere is sandwiched between two stable photon spheres, as indicated in figure \ref{figure2}.

\textbf{Proof of the theorem $n_{\text{stable}} - n_{\text{unstable}} = -1$ in reference \cite{Cunha2020}:} To provide a new proof of the theorem $n_{\text{stable}} - n_{\text{unstable}} = -1$ for black hole spacetimes proposed by Cunha \emph{et al.} in reference \cite{Cunha2020}, we should further demonstrate the innermost photon sphere (or outermost photon sphere) in black hole spacetimes to be an unstable photon sphere. Given the odd total number of photon spheres near black holes $n_{\text{stable}} + n_{\text{unstable}} = 2k+1$ (which have been explained at the end of section \ref{section3}), and the one-to-one alternative separation of stable and unstable photon spheres, the numbers of stable and unstable photon spheres are constrained to be $n_{\text{stable}} = k$ and $n_{\text{unstable}} = k+1$. This fulfills the proof on relation $n_{\text{stable}} - n_{\text{unstable}} = -1$.

To demonstrate the innermost photon sphere near a black hole to be unstable, we once again apply the Gauss-Bonnet theorem to a similar annular region depicted in figure \ref{figure3}. However, in this case, modifications are made for the inner and outer boundaries of region $D$: the inner boundary circle approaches to the event horizon $r_{\text{min}} \to r_{H}$ and the outer boundary circle is the innermost photon sphere $r_{\text{max}} = r_{\text{innermost}}$. The application of Gauss-Bonnet theorem in such region yields
\begin{equation}
	\int_{D}\mathcal{K}\cdot dS + \int_{\partial D} \kappa_{g} dl = 2 \pi \chi(D) = 0 \ . \nonumber
\end{equation}
The contour integral of geodesic curvature along the outer boundary circle (which is a photon sphere) is vanished, which suggests
\begin{eqnarray}
	\int_{D}\mathcal{K}\cdot dS
	& = & - \int_{C_{r}(r \to r_{H})} \kappa_{g} dl  \nonumber
	\\
	& = & - \int_{\phi = 0}^{\phi = -2\pi} \kappa_{g}(r) \sqrt{\tilde{g}_{\phi\phi}^{\text{OP}}} d\phi   \nonumber
	\\
	& = & C(r_{H}) \cdot \lim_{r \to r_{H}} \kappa_{g}(r) \nonumber
	\\
    & \propto & - 2\pi r_{H} \cdot \kappa_{\text{surface}} < 0 \ . 
\end{eqnarray}
Here, $C(r_{H})$ is the circumference of inner boundary $r \to r_{H}$ in 2-dimensional optical geometry. Notably, the relationship between geodesic curvature $\kappa_{g}(r)$ in the near horizon limit and the surface gravity in black hole spacetime $\lim_{r \to r_{H}}\kappa_{g}(r) = \kappa_{\text{surface}} < 0$ is used. Besides, the direction of inner boundary of region $D$ in figure \ref{figure3} is chosen opposite to the increasing of azimuthal angle $\phi$, which also contributes to an additional minus sign.

\begin{figure*}
	\includegraphics[width=0.575\textwidth]{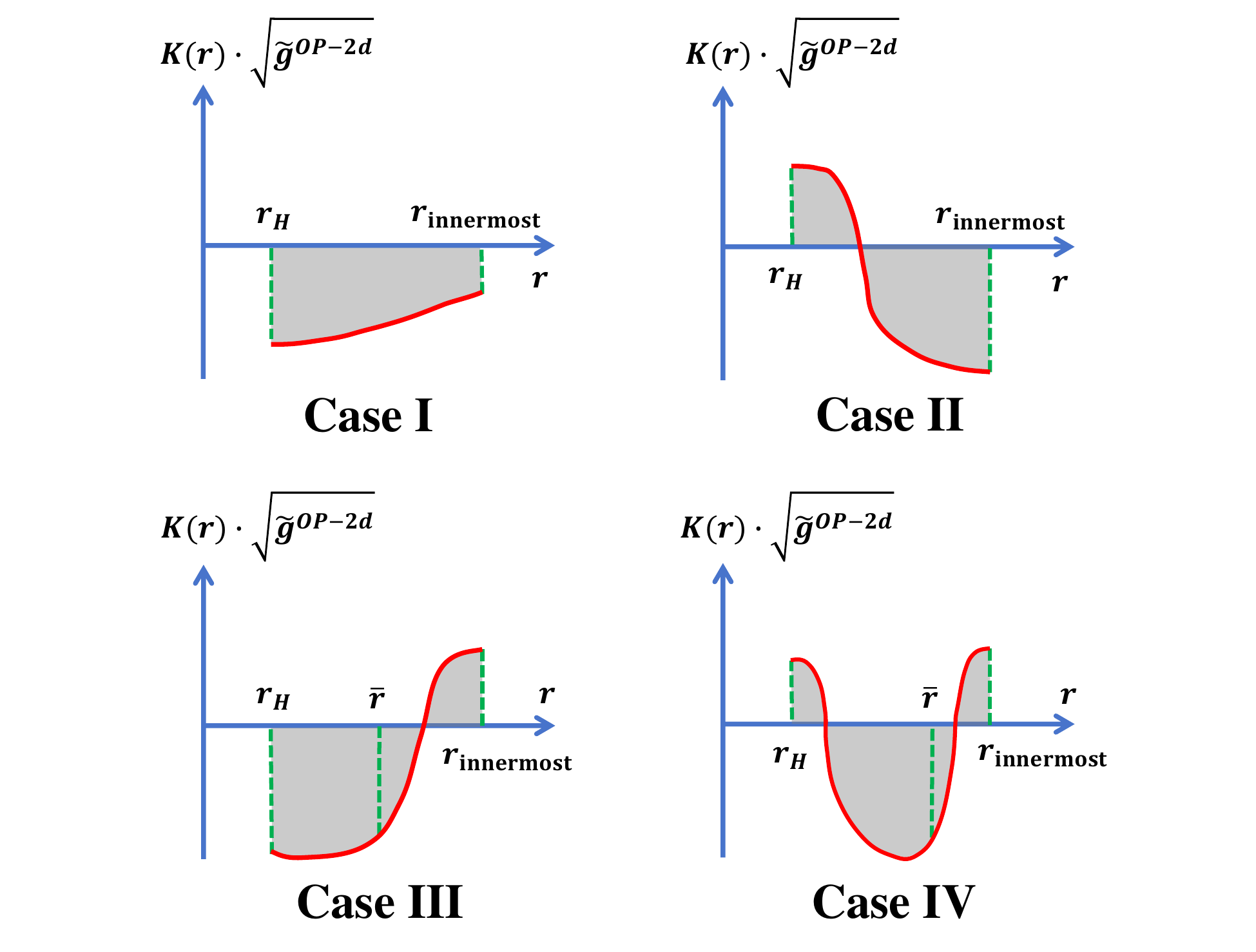}
	\caption{This figure summarizes the four different cases in which the surface integral of Gaussian curvature satisfies $\int_{D}\mathcal{K}\cdot dS = - 2 \pi \kappa_{\text{surface}} < 0$. \textbf{Case I:} the innermost photon sphere is unstable, with $\mathcal{K}(r_{H}) < 0$ and $\mathcal{K}(r_{\text{innermost}}) < 0$. \textbf{Case II:} the innermost photon sphere is unstable, with $\mathcal{K}(r_{H}) > 0$ and $\mathcal{K}(r_{\text{innermost}}) < 0$. \textbf{Case III:} the innermost photon sphere is stable, with $\mathcal{K}(r_{H}) < 0$ and $\mathcal{K}(r_{\text{innermost}}) > 0$. \textbf{Case IV:} the innermost photon sphere is stable, with $\mathcal{K}(r_{H}) > 0$ and $\mathcal{K}(r_{\text{innermost}}) > 0$. In the last two cases, an additional photon sphere $r=\bar{r}$ must exist between the event horizon $r=r_{H}$ and the faked ``innermost'' photon sphere $r=r_{\text{innermost}}$. The factor $\sqrt{\tilde{g}^{\text{OP-2d}}}$ in surface integral is always positive, so the sign of Gaussian curvature can be easily observed from in this figure. This figure shows the situations where Gaussian curvature is continuous in region $D$.}
	\label{figure6}
\end{figure*}

\begin{figure*}
	\includegraphics[width=0.575\textwidth]{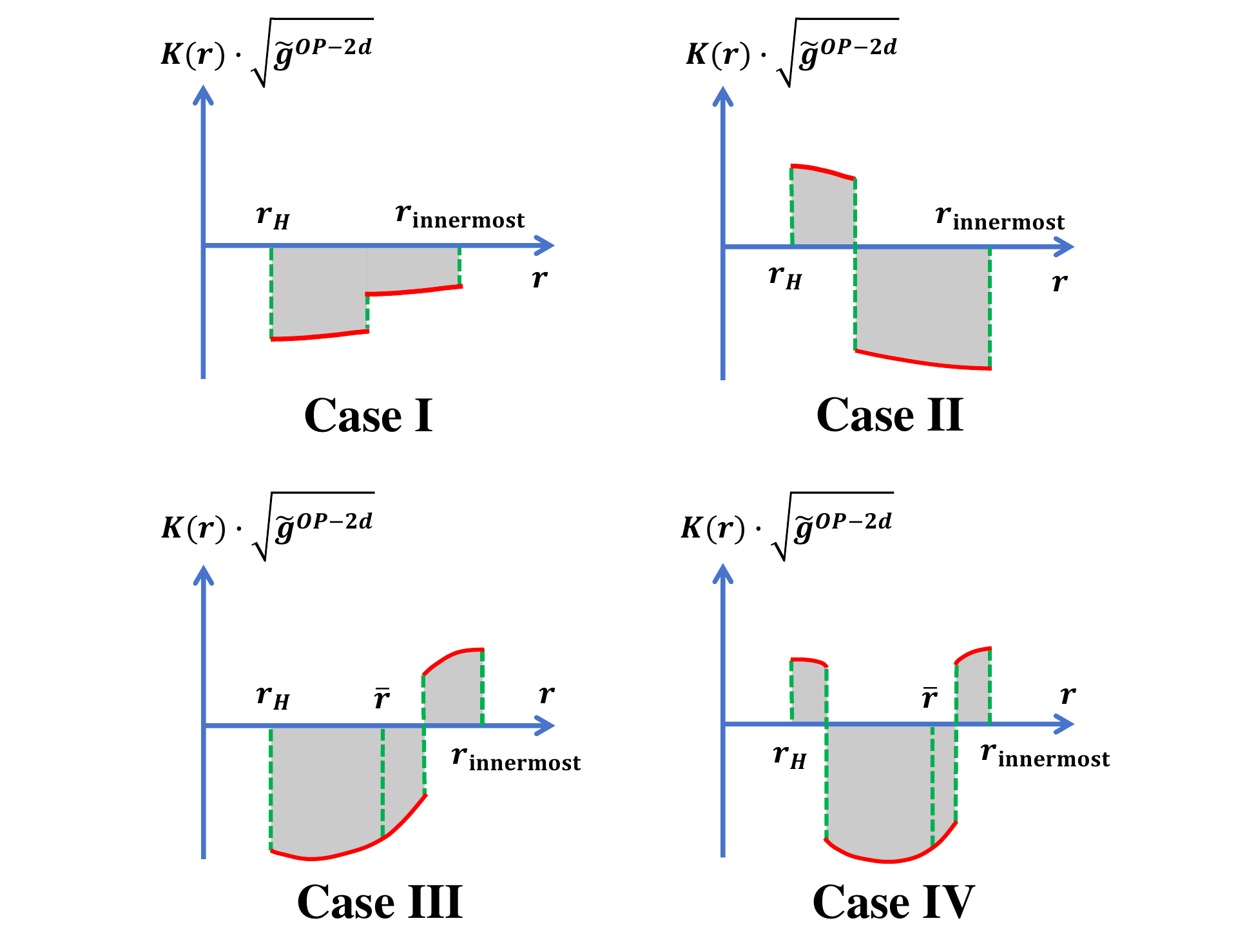}
	\caption{This figure summarizes the four different cases in which the surface integral of Gaussian curvature satisfies $\int_{D}\mathcal{K}\cdot dS = - 2 \pi \kappa_{\text{surface}} < 0$. \textbf{Case I:} the innermost photon sphere is unstable, with $\mathcal{K}(r_{H}) < 0$ and $\mathcal{K}(r_{\text{innermost}}) < 0$. \textbf{Case II:} the innermost photon sphere is unstable, with $\mathcal{K}(r_{H}) > 0$ and $\mathcal{K}(r_{\text{innermost}}) < 0$. \textbf{Case III:} the innermost photon sphere is stable, with $\mathcal{K}(r_{H}) < 0$ and $\mathcal{K}(r_{\text{innermost}}) > 0$. \textbf{Case IV:} the innermost photon sphere is stable, with $\mathcal{K}(r_{H}) > 0$ and $\mathcal{K}(r_{\text{innermost}}) > 0$. In the last two cases, an additional photon sphere $r=\bar{r}$ must exist between the event horizon $r=r_{H}$ and the faked ``innermost'' photon sphere $r=r_{\text{innermost}}$. The factor $\sqrt{\tilde{g}^{\text{OP-2d}}}$ in surface element is always positive, so the sign of Gaussian curvature can be easily observed in the figure. This figure corresponds to the situations where Gaussian curvature admits finite number of discontinuous points with respect to radial coordinate $r$.}
	\label{figure7}
\end{figure*}

There are four different cases that could lead to the surface integral of Gaussian curvature satisfies $\int_{D}\mathcal{K}\cdot dS \propto - 2 \pi r_{H} \kappa_{\text{surface}} < 0$. These cases have been summarized in figure \ref{figure6}. The negative value of surface integral $\int_{D}\mathcal{K}\cdot dS = 2 \pi \int_{r_{H}}^{r_{\text{innermost}}} \mathcal{K}(r) \sqrt{\tilde{g}^{\text{OP-2d}}(r)} dr < 0$ indicates that the shadow areas depicted in figure \ref{figure6} in the integration is negative. In the first two cases, the innermost photon spheres are both unstable photon spheres with negative Gaussian curvature, which is exactly what we need in the demonstration of $n_{\text{stable}} - n_{\text{unstable}} = -1$. However, in the last two cases, it appears that the innermost photon spheres are stable photon spheres, which could hinder the demonstration of $n_{\text{stable}} - n_{\text{unstable}} = -1$. Fortunately, a simple analysis suggests that an additional photon sphere must exist between the event horizon $r=r_{H}$ and the photon sphere $r=r_{\text{innermost}}$. In these cases, the truly innermost photon sphere should locate at a smaller radius $r=\bar{r}$, rather than $r=r_{\text{innermost}}$. As shown in the lower panel of figure \ref{figure6}, in the last two cases, it is always possible to find a new position $r=\bar{r}$ in the interval $[r_{H}, r_{\text{innermost}}]$, such that the integral $\int \mathcal{K}(r) \sqrt{\tilde{g}^{\text{OP-2d}}(r)} dr$ vanishes in interval $[\bar{r}, r_{\text{innermost}}]$ (and the shadow areas in this interval in the lower panel of figure \ref{figure6} are zero). By utilizing the simplified integration formulas in equations (\ref{Gauss-Bonnet reduce}), the vanishing of integral suggests that $H(\bar{r}) = H(r_{\text{innermost}}) = 0$ and $\kappa_{g}(\bar{r}) = \kappa_{g}(r_{\text{innermost}}) = 0$, indicating that an additional photon sphere must exist at position $r=\bar{r}$. Moreover, from the lower panel of figure \ref{figure6}, we observe that the negative Gaussian curvature constrains this new photon sphere at position $r=\bar{r}$ to be an unstable photon sphere. In a summary, in all four cases, we have demonstrated the innermost photon spheres must be unstable, accomplishing the proof of the theorem $n_{\text{stable}} - n_{\text{unstable}} = -1$ in reference \cite{Cunha2020}. Similar to the analysis of photon sphere distribution on previous pages, our discussion presented here still holds even if the Gaussian curvature admits finite discontinuous points, as presented in figure \ref{figure7} \cite{footnote Gauss-Bonnet discontinuity}.

An equivalent way to demonstrate the theorem $n_{\text{stable}} - n_{\text{unstable}} = -1$ is by showing the outermost photon sphere to be an unstable one. This can also be achieved through similar processes using the Gauss-Bonnet theorem for an annular region in figure \ref{figure3}, with the inner boundary circle to be the outermost photon sphere $r_{\text{min}} = r_{\text{outermost}}$ and the outer boundary circle approaching to infinity $r_{\text{max}} \to \infty$. The detailed proof along this direction could be a bit more difficult than what we have presented in this work, as it relies on the detailed asymptotic behavior / asymptotically analytical expansion of the Gaussian curvature in the infinite distance limit.

\section{Summary and Prospects \label{section5}}

The photon sphere / circular photon orbit has become a significantly important topic in black hole physics. In this study, we provide a geometric analysis of photon spheres in spherically symmetric black hole spacetimes. The existence and distribution characteristics of photon spheres in the vicinity of black holes are derived from geometric perspective. Our analysis follows the geometric approach in our recent works, in which the location of photon spheres and their stability are completely determined by geodesic curvature and Gaussian curvature in the optical geometry of black hole spacetimes. 

In the current work, some basic conclusions on photon spheres in black hole spacetime have been rederived. Firstly, for black holes with most commonly asymptotic properties (the asymptotically flat black holes, asymptotically de-Sitter and asymptotically anti de-Sitter black holes), the existence of photon spheres is successfully demonstrated from the behaviors of geodesic curvature in near horizon limit and infinite distance limit. Secondly, applying the Gauss-Bonnet theorem in optical geometry, we demonstrate that the stable photon spheres and unstable photon spheres must be one-to-one alternatively separated from each other (each stable photon sphere is sandwiched between two nearby unstable photon spheres; each unstable photon sphere is sandwiched between two nearby stable photon spheres). Finally, we provide a new geometric proof of the theorem $n_{\text{stable}}-n_{\text{unstble}}=-1$ for black hole spacetimes. Furthermore, our geometric analysis presented in this work offer a new pathway to explore photon spheres and their distributions in gravitational fields, complementing the existing methods (such as the conventional effective potential approach, the topological approach). 

Our analysis in this study is applicable to any spherically symmetric black holes with the general spacetime metric $ds^{2} = g_{tt} dt^{2} + g_{rr} dr^{2} + g_{\theta\theta} d\theta^{2} + g_{\phi\phi} d\phi^{2}$, regardless of the specific analytical expression for metric components. Consequently, a similar analysis can be easily extended to photon spheres in other kinds of gravitational systems, including spacetimes generated by astrophysical ultracompact objects (such as dwarfs and neutron stars) without the presence of event horizons, regular spacetimes in void of singularities, and the naked singularity spacetimes. The detailed and comprehensive analysis of these cases can be carried out in future studies. Hopefully, the geometric analysis provided in current work could be generalized to other gravitational systems, providing us insights and understanding on photon spheres, gravity theories and mathematical physics.

\begin{acknowledgments}
	We would like to thank Ming Li and Yang Huang for valuable discussions. This work is supported by the Scientific and Technological Research Program of Chongqing Municipal Education Commission (Grant No. KJQN202201126), the Natural Science Foundation of Chongqing Municipality (Grant No. CSTB2022NSCQ-MSX0932), the Scientific Research Program of Chongqing Science and Technology Commission (the Chongqing ``zhitongche'' program for doctors, Grant No. CSTB2022BSXM-JCX0100), the Scientific Research Foundation of Chongqing
	University of Technology (Grant No. 2020ZDZ027), and the Research and Innovation Team Cultivation Program of Chongqing University of Technology (Grant No. 2023TDZ007). 
\end{acknowledgments}


\appendix

\section*{Appendix}

The appendices give some mathematical preliminaries needed in the current work. Firstly, the optical geometry of black hole spacetime is reviewed in Appendix \ref{appendix1}. The Gaussian curvature and geodesic curvature in surface theory and differential geometry are presented in Appendix \ref{appendix2}. The Appendix \ref{appendix3} introduces the Gauss-Bonnet theorem in differential geometry. Finally, the relationship between Gaussian curvature and stability of photon spheres (through the Cartan-Hadamard theorem) is provided in Appendix \ref{appendix4}

\section{Optical Geometry of Black Hole Spacetime \label{appendix1}}

The appendix gives an introduction of the optical geometry of black hole spacetime, in which our geometric analysis to photon sphere is carried out. The optical geometry serves as a powerful tool to investigate the motions of photons (or other massless particles that travel along null geodesics) in gravitational fields. The underlying physical interpretation of the mathematical construction of optical geometry can be regarded as a generalization of the Fermat's principle in curved spacetime. 

There are several equivalent ways to construct the optical geometry of black hole spacetimes. Straightforwardly specking, the optical geometry can be obtained from spacetime geometry $ds^{2}=g_{\mu\nu}dx^{\mu}dx^{\nu}$, which is a 4-dimensional Lorentz manifold, through a continuous mapping with the null constraint $d\tau^{2}=-ds^{2}=0$ imposed \cite{Gibbons2008,Ishihara2016a}. 
\begin{equation}
	\underbrace{ds^{2} = g_{\mu\nu}dx^{\mu}dx^{\nu}}_{\text{Spacetime Geometry}}
	\ \ \overset{d\tau^{2}=-ds^{2}=0}{\Longrightarrow} \ \ 
	\underbrace{dt^{2} = g^{\text{OP}}_{ij}dx^{i}dx^{j}}_{\text{Optical Geometry}}
\end{equation}
Another equivalent way (which is more mathematically rigorous) to define an optical geometry is from the conformal transformation of spacetime geometry, followed by extracting the spatial part metric \cite{Gibbons2009}. For static or stationary black holes, the photon orbits (which travels along lightlike / null geodesics in a 4-dimensional Lorentz manifold), become spatial geodesics when they are transformed into optical geometry. The stationary time coordinate $t$ plays the role of arc-length parameter / spatial distance parameter in optical geometry, which must be minimized along the photon orbits. This conclusion can be directly obtained from the generalization of the Fermat's principle in curved manifolds \cite{footnote-Fermat}. 

Furthermore, if we focus on the particle motions in the equatorial plane, a 2-dimensional optical geometry can be constructed.
\begin{equation}
	\underbrace{dt^{2} = g^{\text{OP}}_{ij}dx^{i}dx^{j}}_{\text{Optical Geometry}}
	\ \ \overset{\theta=\pi/2}{\Longrightarrow} \ \ 
	\underbrace{dt^{2}=\tilde{g}^{\text{OP-2d}}_{ij}dx^{i}dx^{j}}_{\text{Optical Geometry (Two Dimensional)}}
\end{equation}
In our geometric analysis, the determination and calculations on photon spheres / circular photon orbits are implemented in this 2-dimensional optical geometry.

The properties of optical geometry strongly depend on the symmetries of the gravitational field and black hole spacetime. For a spherically symmetric black hole, its optical geometry gives rise to a Riemannian manifold \cite{Abramowicz1988,Gibbons2008,Gibbons2009}. For a rotational / axi-symmetric black hole, the corresponding optical geometry is described by a Randers-Finsler manifold \cite{Werner2012,Ono2017,Jusufi2018,Jusufi2018b}. 

Firstly, we present the optical geometry formulation for spherically symmetric black holes. Considering a general spherically symmetric black hole with spacetime metric
\begin{equation}
	ds^{2} = g_{tt}dt^{2}+g_{rr}dr^{2}+g_{\theta\theta}d\theta^{2}+g_{\phi\phi}d\phi^{2} ,
\end{equation} 
the corresponding optical geometry can be obtained from the null constraint $d\tau^{2}=-ds^{2}=0$, which eventually gives a 3-dimensional Riemannian manifold
\begin{equation}
	dt^{2} = g_{ij}^{\text{OP}}dx^{i}dx^{j}
	= -\frac{g_{rr}}{g_{tt}}\cdot dr^{2} - \frac{g_{\theta\theta}}{g_{tt}} \cdot d\theta^{2} - \frac{g_{\phi\phi}}{g_{tt}} d\phi^{2} .
	\label{optical geometry3}
\end{equation}
When analyzing the photon spheres in the vicinity of spherically symmetric black holes, one can always restrict this optical geometry to the equatorial plane $\theta=\pi/2$ without loss of generality. The explicit form of the 2-dimensional optical geometry is
\begin{equation}
	dt^{2}=\tilde{g}^{\text{OP-2d}}_{ij}dx^{i}dx^{j}
	= - \frac{g_{rr}}{g_{tt}} \cdot dr^{2} - \frac{\overline{g}_{\phi\phi}}{g_{tt}} \cdot d\phi^{2} , 
\end{equation}
where $\tilde{g}^{\text{OP-2d}}_{ij}$ denotes the 2-dimensional optical geometry metric, and the simplified notation $\overline{g}_{\phi\phi}$ is defined as $\overline{g}_{\phi\phi}=g_{\phi\phi}(\theta=\pi/2)$. 

Then we give the optical geometry of rotational / axi-symmetric black holes, which is a Randers-Finsler manifold. Considering the standard rotational black hole spacetime metric
\begin{equation}
	ds^{2} = g_{tt}dt^{2}+2g_{t\phi}dtd\phi+g_{rr}dr^{2}+g_{\theta\theta}d\theta^{2}+g_{\phi\phi}d\phi^{2} ,
\end{equation} 
the optical geometry can be obtained in a similar way by imposing the null constraint $d\tau^{2}=-ds^{2}=0$. Eventually, the arc length parameter / spatial distance parameter in optical geometry becomes
\begin{eqnarray}
	dt & = & \sqrt{-\frac{g_{rr}}{g_{tt}} \cdot dr^{2} - \frac{g_{\theta\theta}}{g_{tt}} \cdot d\theta^{2} + \frac{g_{t\phi}^{2}-g_{tt}g_{\phi\phi}}{g_{tt}^{2}} \cdot d\phi^{2}} \nonumber
	\\
	&   & - \frac{g_{t\phi}}{g_{tt}}\cdot d\phi ,
\end{eqnarray}
which exactly gives a Renders-Finsler manifold. Mathematically, the Renders-Finsler geometry is an extension of the Riemannian geometry \cite{ChernSS}, allowing the separation of arc-length / spatial distance into two parts
\begin{equation}
	dt = \sqrt{\alpha_{ij}dx^{i}dx^{j}} + \beta_{i}dx^{i} . \label{Renders geometry}
\end{equation}
The first part $\alpha_{ij}$ is a Riemannian metric, and the second part $\beta=\beta_{i}dx^{i}$ is a one-form that quantifies the departure of this Renders-Finsler geometry from the Riemannian geometry $dt^{2}=\alpha_{ij}dx^{i}dx^{j}$. The above Renders-Finsler geometry recovers the Riemannian geometry if and only if $\beta=0$.

In this work, we only focus on spherically symmetric black holes, whose optical geometry is a Riemannian manifold. In the 2-dimensional Riemannian manifold, the most important intrinsic curvatures are Gaussian curvature and geodesic curvature, with the detailed introductions provided in Appendix \ref{appendix3}.

\section{Gaussian Curvature and Geodesic Curvature \label{appendix2}}

There are several important quantities which describe the geometry of optical geometry. In our geometric analysis, the Gaussian curvature and geodesic curvature play central roles in determining the location of photon spheres and their stability. This appendix gives a brief introduction on the Gaussian curvature and geodesic curvature in differential geometry. 

In the surface theory and differential geometry, the Gaussian curvature is the intrinsic curvature of a 2-dimensional surface $S$, which measures how much this surface deviates from being flat intrinsically. The geodesic curvature is the curvature of a continuous curve $\gamma(s)$ residing on this surface, which measures how far this curve departs from being a geodesic on surface $S$ \cite{Carmo1976,Berger,Berger2}.  If the curve is a geodesic curve on surface $S$, its geodesic curvature automatically vanishes ($\kappa_{g}(\gamma)=0$). Both geodesic curvature and Gaussian curvature are intrinsic geometric quantities, and they can be calculated purely from the metric of the 2-dimensional surface $S$, regardless of the embedding of this surface in a higher dimensional space / spacetime. 

To give the expressions of Gaussian curvature and geodesic curvature, we assign the curving linear coordinates $(x_{1},x_{2})$ for the two-dimensional surface $S$ such that its intrinsic metric for this surface can be expressed as
\begin{equation}
	ds^{2} = g_{11} \cdot dx_{1}^{2} + g_{22} \cdot dx_{2}^{2} \ . \label{intrinsic metric}
\end{equation}
For an arbitrary curve $\gamma=\gamma(s)=(x_{1}(s),x_{2}(s))$ that resides on this surface with arc-length parameter / spatial distance parameter $s$, its geodesic curvature can be calculated through the Liouville's relation \cite{Carmo1976,ChernWH}
\begin{eqnarray}
	\kappa_{g}(\gamma) & = & \frac{d\alpha}{ds} 
	- \frac{1}{2\sqrt{g_{22}}}\frac{\partial \ \text{log}(g_{11})}{\partial x_{2}} \cos\alpha \nonumber
	\\
	&   & + \frac{1}{2\sqrt{g_{11}}}\frac{\partial \ \text{log}(g_{22})}{\partial x_{1}} \sin\alpha \ ,
	\label{geodesic-Curvature1}
\end{eqnarray}
where $\alpha$ is the angle between tangent vector of $\gamma(s)$ and the first coordinate axis $x_{1}$. Since the geodesic curvature quantifies how much a curve deviates from being a geodesic curve on a 2-dimensional surface $S$, the vanishing of geodesic curvature indicates that the corresponding orbit $\gamma=\gamma(s)$ is a geodesic curve on this surface.
\begin{equation}
	\kappa_{g}(\gamma) = 0 
	\Leftrightarrow
	\nabla_{T} T = 0 
	\Leftrightarrow 
	\bigg[ \frac{d^{2}x_{i}}{ds^{2}}-\Gamma_{jk}^{i}\frac{dx_{j}}{ds}\frac{dx_{k}}{ds} \bigg]_{\gamma(s)} = 0 \ \ \ \ 
\end{equation}
Here, $T$ represents the tangent vector for this curve $\gamma$ on a two-dimensional surface.  

Furthermore, the Gaussian curvature of the 2-dimensional surface $S$ can also be calculated using the intrinsic metric in equation (\ref{intrinsic metric}), the detailed expression of Gaussian curvature gives \cite{Carmo1976,ChernWH}
\begin{eqnarray}
	\mathcal{K} & = & \mathcal{K}_{1} \cdot \mathcal{K}_{2}
	= \frac{R_{1212}}{g_{11} \cdot g_{22}-(g_{12}^{2})} \nonumber
	\\
	& = & 
	-\frac{1}{\sqrt{\text{det} g}}
	\bigg[
	\frac{\partial}{\partial x_{2}} \bigg( \frac{1}{\sqrt{g_{22}}} \frac{\partial\sqrt{g_{11}}}{\partial x_{2}}  \bigg)
	+ \frac{\partial}{\partial x_{1}} \bigg( \frac{1}{\sqrt{g_{11}}} \frac{\partial\sqrt{g_{22}}}{\partial x_{1}}  \bigg)
	\bigg] , \nonumber
	\\ \label{Gauss-Curvature1}
\end{eqnarray}
where $\text{det} g =g_{11} \cdot g_{22}$ is the determinant of the 2-dimensional intrinsic metric of $S$ in equation (\ref{intrinsic metric}), the $\mathcal{K}_{1}$ and $\mathcal{K}_{2}$ are principal curvatures of surface $S$ (refer to the Supplemental Material), and $R_{1212}$ denotes the Riemannian curvature tensor of this surface. Notably, in this appendix, we adopt the notations in surface theory, where the covariant and contravariant coordinates are not distinguished.

\section{Gauss-Bonnet Theorem \label{appendix3}}

\begin{figure}
	\includegraphics[width=0.5\textwidth]{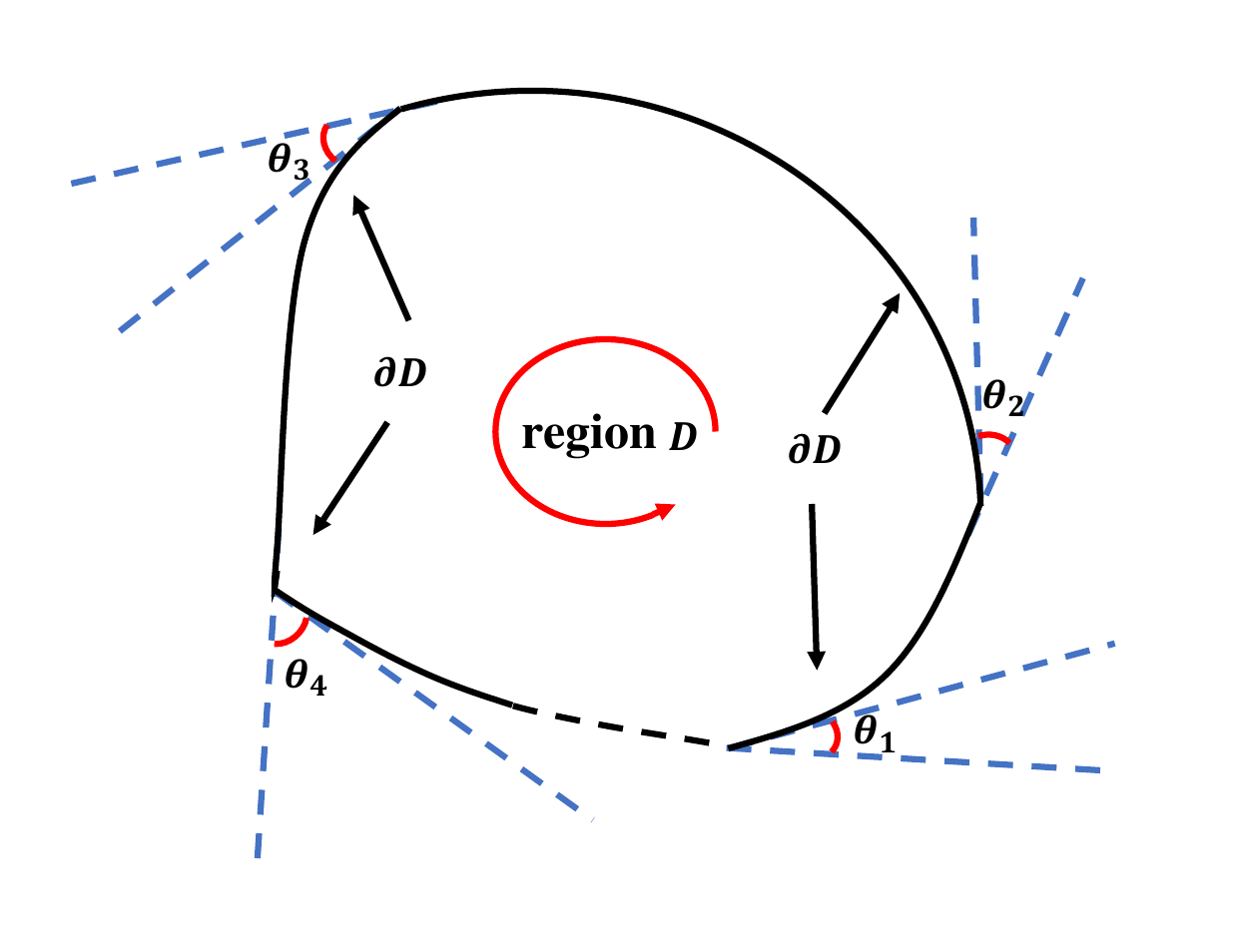}
	\caption{The Gauss-Bonnet theorem for a region $D$ in 2-dimensional manifold.}
	\label{figure Gauss-Bonnet}
\end{figure}

The Gauss-Bonnet theorem is a famous achievement in differential geometry and geometrical topology. It reveals the nontrivial relationship between the local curvatures and the global topological invariant of a curved manifold.   
The mathematical description of Gauss-Bonnet theorem in a two-dimensional curved manifold is
\begin{equation}
	\int_{D} \mathcal{K} dS + \int_{\partial D} \kappa_{g} dl + \sum_{i=1}^{N} \theta_{i} = 2 \pi \chi(D) \ .
	\label{Gauss-Bonnet theorem}
\end{equation} 
In the expression, $D$ is a selected region on a 2-dimensional curved surface, $\mathcal{K}$ labels the Gaussian curvature of this surface, $\kappa_{g}$ is the geodesic curvature of the boundary $\partial D$, $\chi(D)$ denotes the Euler characteristic number for region $D$, and $\theta_{i}$ represents the exterior angle for each discontinuous point on the boundary $\partial D$. The illustration of Gauss-Bonnet theorem in a 2-dimensional manifold is shown in figure \ref{figure Gauss-Bonnet}.

Furthermore, a recent work stated that the surface integral of Gaussian curvature in the optical geometry / Jacobi geometry can be effectively simplified \cite{Huang2022,Huang2023}. In order to present this simplification process, we choose the case of optical geometry as an example. For spherically symmetric black holes, from the explicit formula for Gaussian curvature in the 2-dimensional optical geometry in equation (\ref{Gauss cuurvature expression}), the surface integral for Gaussian curvature over a given region can be reduced to
\begin{eqnarray}
	&   & \int_{D}\mathcal{K} \cdot dS \nonumber
	\\
	& = & \int_{D}\mathcal{K} \cdot \sqrt{\tilde{g}^{\text{OP-2d}}} \cdot dr d\phi
	\nonumber
	\\ 
	& = & 
	\int_{\phi_{\text{min}}}^{\phi_{\text{max}}} d\phi 
	\int_{r_{\text{min}}(\phi)}^{r_{\text{max}}(\phi)} \mathcal{K}(r) 
	\cdot \sqrt{\tilde{g}^{\text{OP-2d}}(r)} \cdot dr \nonumber
	\\
	& = & 
	- \int_{\phi_{\text{min}}}^{\phi_{\text{max}}} d\phi 
	\int_{r_{\text{min}}(\phi)}^{r_{\text{max}}(\phi)}
	\bigg[
	\frac{\partial}{\partial \phi} \bigg( \frac{1}{\sqrt{\tilde{g}^{\text{OP-2d}}_{\phi\phi}}} \frac{\partial\sqrt{\tilde{g}^{\text{OP-2d}}_{rr}}}{\partial \phi}  \bigg) \nonumber
	\\
	&   & \ \ \ \ \ \ \ \ \ \ \ \ \ \ \ \ \ \ \ \ \ \ \ \ \ \ 
	+ \frac{\partial}{\partial r} \bigg( \frac{1}{\sqrt{\tilde{g}^{\text{OP-2d}}_{rr}}} \frac{\partial\sqrt{\tilde{g}^{\text{OP-2d}}_{\phi\phi}}}{\partial r}  \bigg)
	\bigg] dr
	\nonumber
	\\
	& = & 
    - \int_{\phi_{\text{min}}}^{\phi_{\text{max}}} d\phi 
    \int_{r_{\text{min}}(\phi)}^{r_{\text{max}}(\phi)}
    \frac{\partial}{\partial r} \bigg( \frac{1}{\sqrt{\tilde{g}^{\text{OP-2d}}_{rr}}} \frac{\partial\sqrt{\tilde{g}^{\text{OP-2d}}_{\phi\phi}}}{\partial r}  \bigg)
    \cdot dr 
    \label{simplified surface integral}
    \nonumber
    \\
\end{eqnarray}
The metric components of a 2-dimensional optical geometry for spherically symmetric black holes are independent of the azimuthal angle $\phi$, and their partial derivative with respect to $\phi$ vanishes. It is worth noting that the integration function in (\ref{simplified surface integral}) becomes a total derivative, allowing us to defined a primitive function
\begin{eqnarray}
	H(r) & = & - \frac{1}{\sqrt{\tilde{g}^{\text{OP-2d}}_{rr}}} \frac{\partial\sqrt{\tilde{g}^{\text{OP-2d}}_{\phi\phi}}}{\partial r} \nonumber
	\\
	& = & - \frac{1}{2\sqrt{\tilde{g}^{\text{OP-2d}}}}
	\cdot \frac{\partial \tilde{g}^{\text{OP-2d}}_{\phi\phi}}{\partial r} \nonumber
	\\
	& = & - \sqrt{\tilde{g}^{\text{OP-2d}}_{\phi\phi}} \cdot \kappa_{g}(r)
\end{eqnarray} 
with $\tilde{g}^{\text{OP-2d}} = \text{det} (\tilde{g}^{\text{OP-2d}}) = \tilde{g}^{\text{OP-2d}}_{rr} \cdot \tilde{g}^{\text{OP-2d}}_{\phi\phi}$ to be the determinant of 2-dimensional optical metric. With the primitive function $H(r)$, we eventually obtain the simplified expression for surface integral given in references \cite{Huang2022,Huang2023}
\begin{equation}
	\int_{D}\mathcal{K} \cdot dS
	=
	\int_{\phi_{\text{min}}}^{\phi_{\text{max}}} 
	\big[ H(r_{\text{max}}(\phi)) - H(r_{\text{min}}(\phi)) \big] \cdot d\phi 
\end{equation}
Particularly, for the selected annular region in figure \ref{figure3}, both the minimal and maximal radius $r_{\text{min}}(\phi)$, $r_{\text{max}}(\phi)$ in region $D$ are photon sphere radius independent of azimuthal angle. Therefore, we deduce the simplified formula (\ref{Gauss-Bonnet reduce}) in the Gauss-Bonnet theorem.

\section{Gaussian Curvature and the Stability of Photon Spheres \label{appendix4}}

\begin{figure*}
	\includegraphics[width=0.495\textwidth]{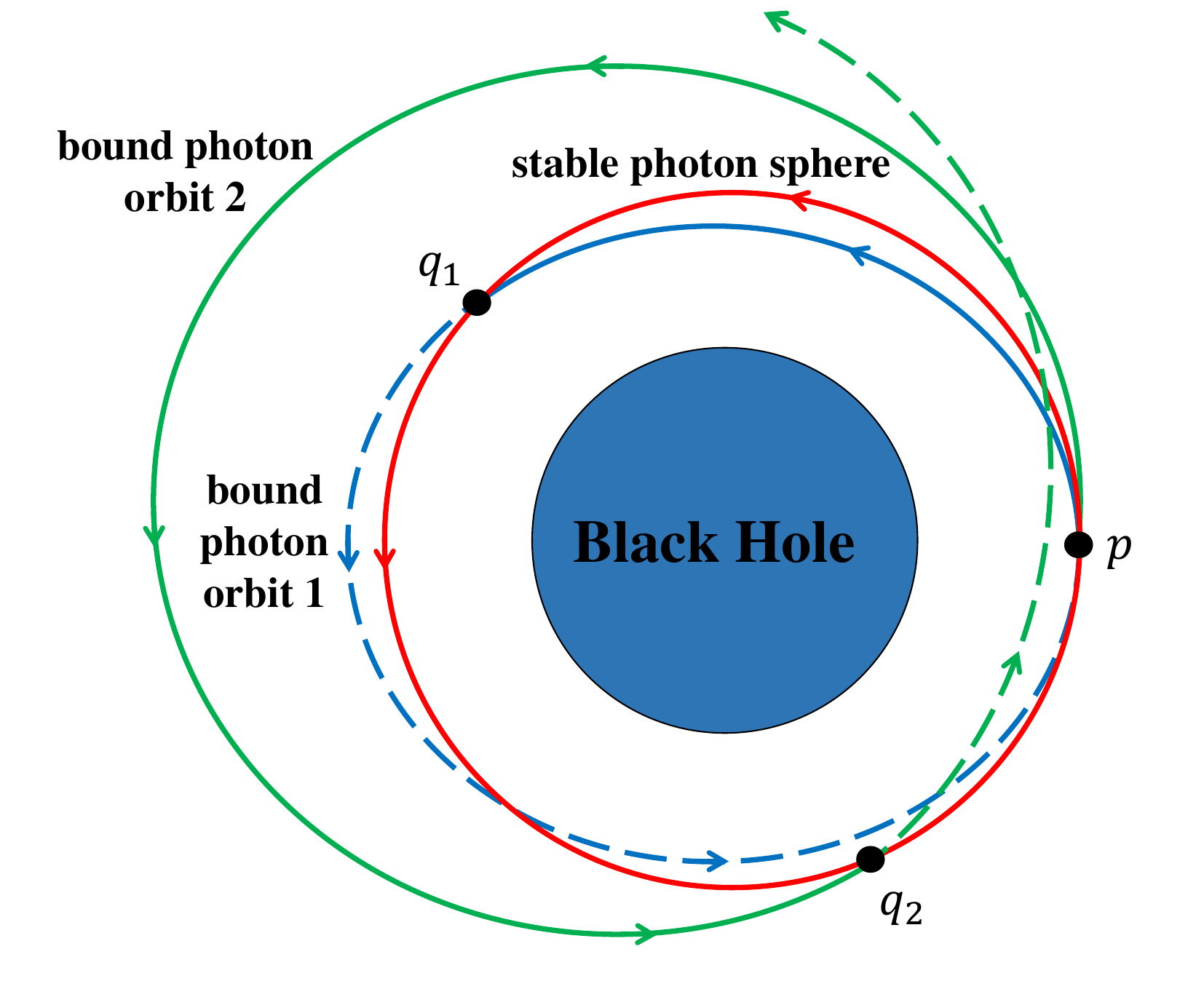}
	\includegraphics[width=0.495\textwidth]{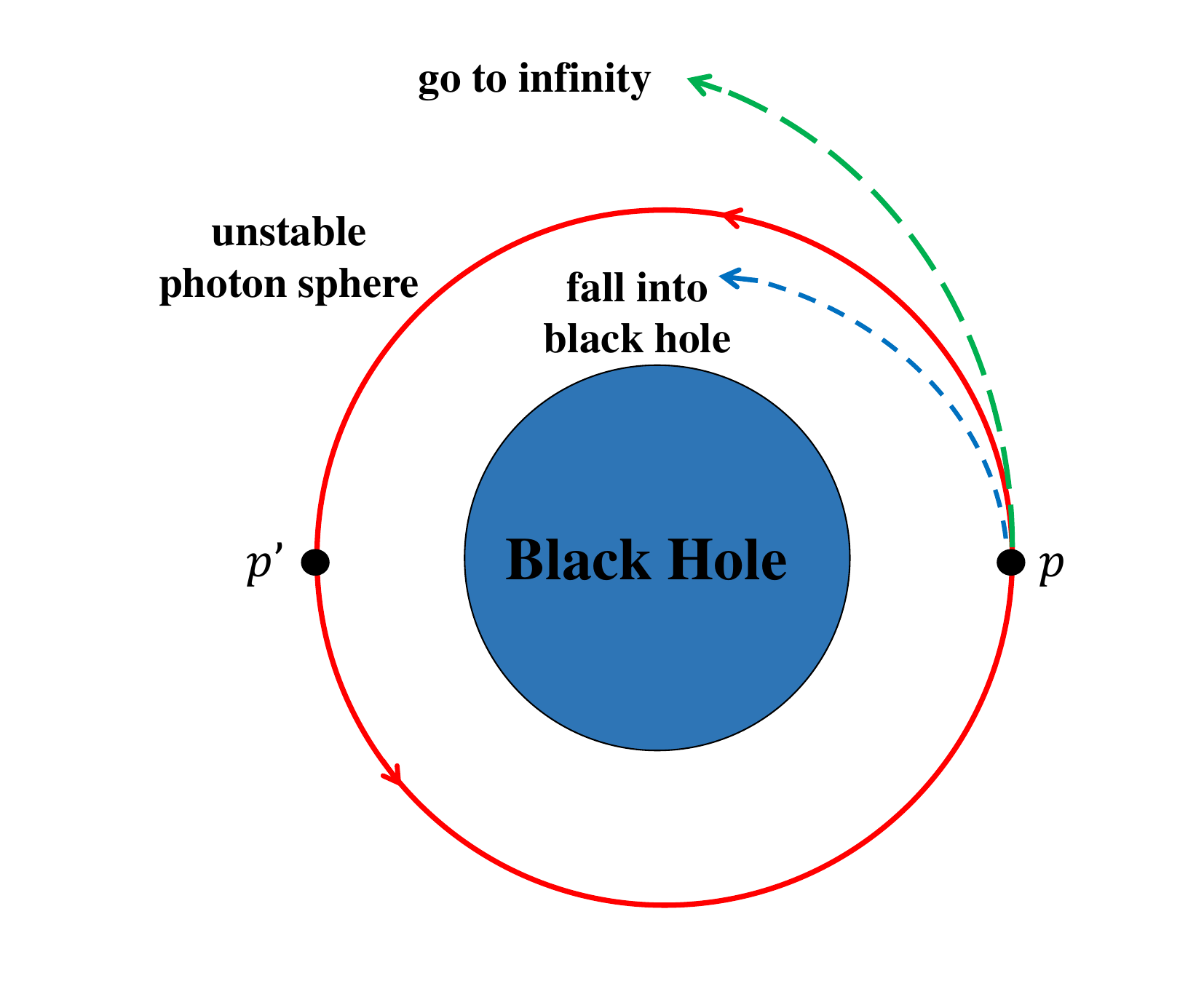}
	\caption{This figure illustrates the possible bound photon orbits near the stable and unstable photon spheres. 
	\textbf{Left panel:} In the cases of stable photon spheres, other bound photon orbits can be obtained from a small perturbation of the photon sphere at a particular point $p$. These bound photon orbits may exhibit different shapes. The first bound photon orbit labeled in blue is a closed orbit, and the second bound photon orbit labeled in green is an unclosed orbit with procession. In the unstble photon sphere, it is always possible to find another point $q$ (which is denote as $q_{i}$ for bound orbits $i=1,2$) conjugated to point $p$. When we consider photon orbits starting from $p$ ending at $q$, there are at least two different geodesic curves (one is the stable photon sphere, and the other is a bound photon orbit) that can be continuously deformed to each other. These two different geodesic curves from $p$ to $q$ belonging to the same homotopy class.
	\textbf{Right panel:} In the cases of unstable photon spheres, the photon beams may either fall into the black black or go to infinity when they are perturbed from the photon sphere at a particular point $p$. It is impossible to find a point $q$ conjugated to point $p$, such that two different geodesic curves from $p$ to $q$ belong to the same homotopy class.}
	\label{figure Photon Sphere}
\end{figure*}

In this appendix, we give a brief introduction to the relationship between Gaussian curvature and the stability of photon spheres near black holes. The descriptions presented here can be viewed as a concise overview of the analysis in references \cite{Qiao2022a,Qiao2022b}. Within our geometric approach, the stability of photon spheres is directly connected with the sign of Gaussian curvature, and the underlying mathematical nature is revealed from the Cartan-Hadamard theorem in differential geometry and topology. Specifically, the negative Gaussian curvature indicates the corresponding photon spheres are unstable, while the positive Gaussian curvature implies the corresponding photon spheres are stable.

In the gravitational field, the photon spheres in the vicinity of black holes and other ultra-compact objects can be classified into two categories: stable photon spheres and unstable photon spheres. These two categories of photon spheres could exhibit significantly different features. For unstable photon spheres, when photon beams have a perturbation from the photon sphere at a particular point $p$, they would either fall into the black hole or go to infinity. In other words, no bound photon orbits are admitted near the unstable photon sphere. Conversely, when photon beams have a perturbation from the stable photon sphere at point $p$, they could also moving around in the nearby orbits. In other words, there are other bound photon orbits exist near the stable photon sphere. These bound photon orbits may have different shapes, including the closed orbit or the unclosed orbits with procession (similar to the planetary orbit around the Sun). The schematic representation of stable and unstable photon spheres and the photon motion through a perturbation are illustrated schematically in figure \ref{figure Photon Sphere}. 

\begin{figure}
	\includegraphics[width=0.5\textwidth]{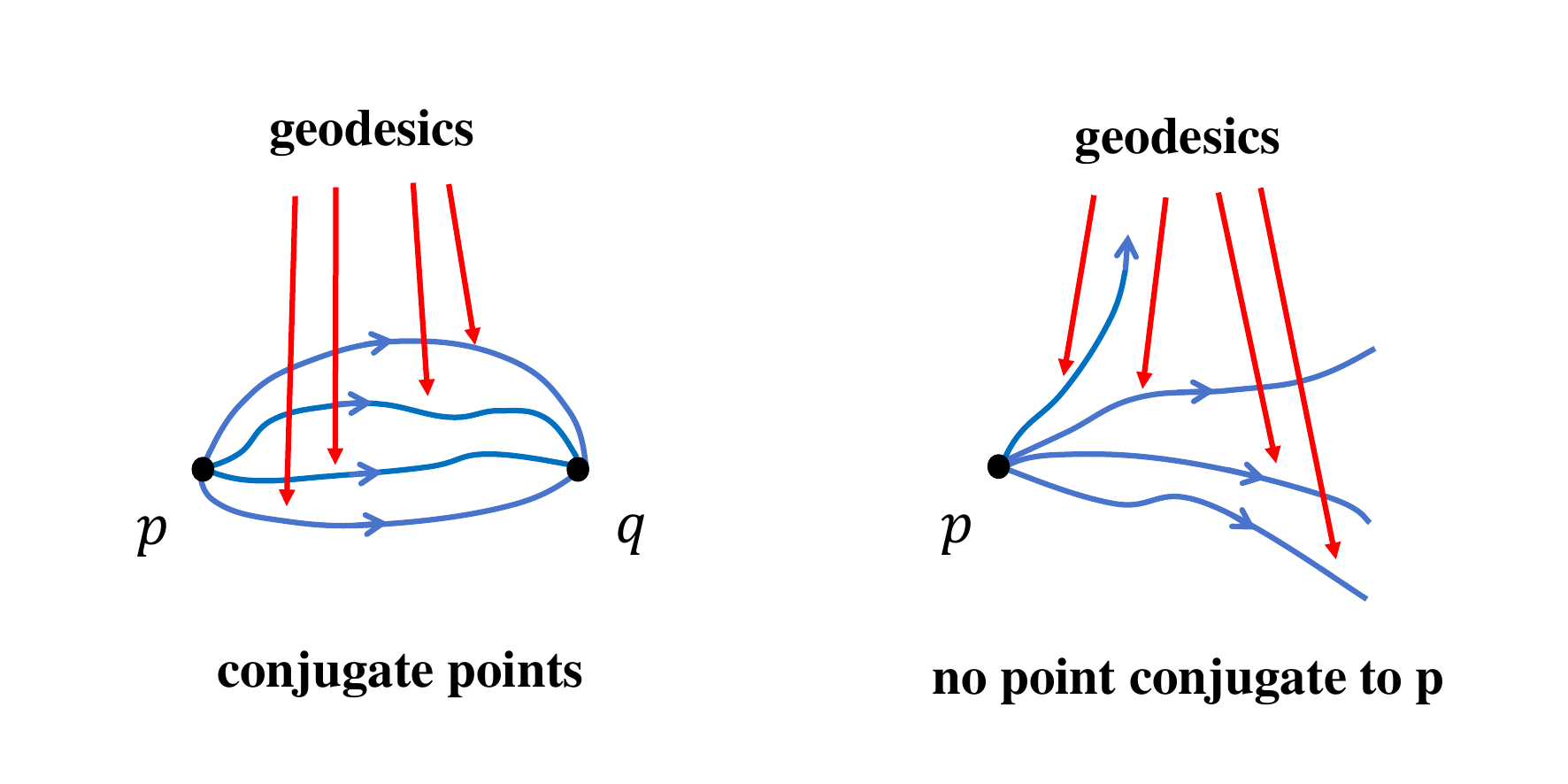}
	\caption{This figure illustrates the mathematical concept of conjugate points in a manifold. The left panel shows that two points $p$ and $q$ are conjugate to each other. The right panel depicts the cases where there are no points in the manifold that conjugate to a given point $p$.}
	\label{figure Conjugate Points}
\end{figure}

Mathematically, these distinct characteristics of stable and unstable photon spheres reflect the nature of conjugate points in a curved manifold. Conjugate points are important concept in modern differential geometry and topology. In a manifold, two points $p$ and $q$ are defined to be conjugate to each other, if and only if there are at least two different geodesic curves starting from point $p$ could eventually converge to point $q$. On the other hand, if a particular point $p$ has no conjugate point in this manifold, then different geodesic curves starting from point $p$ cannot converge to any other points \cite{Berger2}. The figure \ref{figure Conjugate Points} gives a sketch of the conjugate points. Once there exist conjugate points $p$ and $q$ in the manifold, the different geodesics, which start from a given point $p$ and converge to a fixed point $q$, can be continuously transformed to each other through a continuous mapping. Topologically, these geodesics from $p$ to $q$ belong to the same homotopy class, as illustrated in the left panel of figure \ref{figure Conjugate Points}.

Based on the above analysis, since the distinct characteristics of stable and unstable photon spheres are reflected by conjugate points, the existence of conjugate points in the manifold may provide us a new viewpoint to distinguish the stable and unstable photon spheres. Mathematically, the presence of conjugate points can be strongly influenced by the Gaussian curvature. Particularly, the classical Cartan-Hadamard theorem in differential geometry and topology establishes a nontrivial relationship between conjugate points and the Gaussian curvature in a 2-dimensional manifold.
\begin{quote}
	\textbf{Cartan-Hadamard Theorem:}
	For a 2-dimensional complete Riemannian manifold with nonpositive Gaussian curvature, there exists a unique geodesic curve connecting a starting point $p$ to an end point $p'$ belong to the same homotopy class, and this geodesic curve minimizes the length among all curves in this homotopy class. Under such condition, there are no points $q$ that are conjugate to the given point $p$ in the 2-dimensional manifold \cite{Berger2,footnote-geodesic-complete}.
\end{quote} 

Applying the Cartan-Hadamard theorem in the 2-dimensional optical geometry, an analysis on the stability of photon spheres can be conducted. In the cases of stable photon spheres, due to the presence of other bound photon orbits near photon spheres, it is always possible to find conjugate points $p$ and $q$ in the stable photon sphere such that there exist at least two geodesic curves (one is the stable photon sphere, the other is a bound photon orbit) belonging to the same homotopy class. The left panel of figure \ref{figure Photon Sphere} illustrates two possible bound photon orbits and their corresponding conjugate points $p$ and $q$, respectively. In such cases, the Gaussian curvature should be positive, otherwise it would violate the Cartan-Hadamard theorem. On the contrary, in the cases of unstable photon spheres, it is impossible to find any conjugate points $p$ and $q$ in the unstable photon sphere. As shown in the right panel of figure \ref{figure Photon Sphere}, there is a unique geodesic curve (the unstable photon sphere itself) within the entire homotopy class (for continuous curves starting from a given point $p$ ending at an another point $p'$), which exactly corresponds to the negative Gaussian curvature in the Cartan-Hadamard theorem. To summarize, the following criterion can be used to distinguish the stable and unstable photon spheres 
\begin{eqnarray}
	\mathcal{K} < 0 & \Rightarrow & \text{The photon sphere $r=r_{ph}$ is unstable}  \nonumber \\
	\mathcal{K} > 0 & \Rightarrow & \text{The photon sphere $r=r_{ph}$ is stable} \nonumber
\end{eqnarray}




\begin{thebibliography}{99}
	
	\bibitem{ETH2019a}
	K. Akiyama \emph{et al.} (The Event Horizon Telescope Collaboration), First M87 Event Horizon Telescope Results. I. The Shadow of the Supermassive Black Hole, 
	\href{https://doi.org/10.3847/2041-8213/ab0ec7}{Astrophys. J. {\bf 875}, L1 (2019).}  \href{https://arxiv.org/abs/1906.11238}{arXiv:1906.11238 [astro-ph.GA]}
	
	\bibitem{ETH2022}
	K. Akiyama \emph{et al.} (The Event Horizon Telescope Collaboration), First Sagittarius A* Event Horizon Telescope Results. I. The Shadow of the Supermassive Black Hole in the Center of the Milky Way, \href{https://doi.org/10.3847/2041-8213/ac6674}{Astrophys. J. Lett. {\bf 930} 2, L12 (2022).}
	
	\bibitem{Johannsen2013}
	Tim Johannsen, \emph{Photon Rings around Kerr and Kerr-like Black Holes}, 
	\href{https://doi.org/10.1088/0004-637X/777/2/170}{Astrophys. J. {\bf 777}, 170 (2013).}    \href{https://arxiv.org/abs/0805.3146}{arXiv:0805.3146[gr-qc]}
	
	\bibitem{Hod2013}
	S. Hod, Upper bound on the radii of black-hole photonspheres,
	\href{https://doi.org/10.1016/j.physletb.2013.10.047}{Phys. Lett. B \textbf{727}, 345-348 (2013).}
	\href{https://doi.org/10.48550/arXiv.1701.06587}{arXiv:1701.06587[gr-qc]}
	
	\bibitem{Tsukamoto2014}
	N. Tsukamoto, Z. Li and C. Bambi, Constraining the spin and the deformation parameters from the black hole shadow,
	\href{https://doi.org/10.1088/1475-7516/2014/06/043}{J. Cosmol. Astropart. Phys. \textbf{06}, 043 (2014).}
	\href{https://doi.org/10.48550/arXiv.1403.0371}{arXiv:1403.0371[gr-qc]}
	
	\bibitem{Atamurotov2015}
	F. Atamurotov and B. Ahmedov, Optical properties of black hole in the presence of plasma: shadow,
	\href{https://doi.org/10.1103/PhysRevD.92.084005}{Phys. Rev. D \textbf{92}, 084005 (2015).}
	\href{https://doi.org/10.48550/arXiv.1507.08131}{arXiv:1507.08131[gr-qc]}
	
	\bibitem{Tsukamoto2018}
	N.~Tsukamoto, Black hole shadow in an asymptotically-flat, stationary, and axisymmetric spacetime: The Kerr-Newman and rotating regular black holes,
	\href{https://doi.org/10.1103/PhysRevD.97.064021}{Phys. Rev. D \textbf{97}, 064021 (2018).}
	\href{https://doi.org/10.48550/arXiv.1708.07427}{arXiv:1708.07427[gr-qc]}
	
	\bibitem{Perlick2018}
	V. Perlick, O. Y. Tsupko and G. S. Bisnovatyi-Kogan, Black hole shadow in an expanding universe with a cosmological constant,
	\href{https://doi.org/10.1103/PhysRevD.97.104062}{Phys. Rev. D \textbf{97}, 104062 (2018).}
	\href{https://doi.org/10.48550/arXiv.1804.04898}{arXiv:1804.04898[gr-qc]}
	
	\bibitem{JiaJJ2018a}
	J. Jia, J. Liu, X. Liu, Z. Mo, X. Pang, Y. Wang and N. Yang, Existence and stability of circular orbits in general static and spherically symmetric spacetimes, \href{https://doi.org/10.1007/s10714-017-2337-1}{Gen. Rel. Grav. {\bf 50}, 17 (2018).}
	\href{https://doi.org/10.48550/arXiv.1702.05889}{arXiv:1702.05889[gr-qc]}
	
	\bibitem{JiaJJ2018b}
	J. Jia, X. Pang and N. Yang, Existence and stability of circular orbits in static and axisymmetric spacetimes, \href{https://doi.org/10.1007/s10714-018-2364-6}{Gen Relativ Gravit {\bf 50}, 41 (2018).} 
	\href{https://doi.org/10.48550/arXiv.1704.01689}{arXiv:1704.01689[gr-qc]}
	
	\bibitem{Shaikh2018}
	R. Shaikh, P. Kocherlakota, R. Narayan and P. S. Joshi, Shadows of spherically symmetric black holes and naked singularities, \href{https://doi.org/10.1093/mnras/sty2624}{Mon. Not. Roy. Astron. Soc. \textbf{482}, 52-64 (2019).}
	\href{https://doi.org/10.48550/arXiv.1802.08060}{arXiv:1802.08060[astro-ph.HE]}
	
	\bibitem{WeiSW2019}
	S. -W. Wei, Y. -X. Liu and R. B. Mann, Intrinsic curvature and topology of shadows in Kerr spacetime, \href{https://doi.org/10.1103/PhysRevD.99.041303}{Phys. Rev. D 99, 041303(R) (2019).}   \href{https://doi.org/10.48550/arXiv.1811.00047}{arXiv:1811.00047[gr-qc]}
	
	\bibitem{Mishra2019}
	A. K. Mishra, S. Chakraborty and S. Sarkar, Understanding photon sphere and black hole shadow in dynamically evolving spacetimes, \href{https://doi.org/10.1103/PhysRevD.99.104080}{Phys. Rev. D \textbf{99}, 104080 (2019).}
	\href{https://doi.org/10.48550/arXiv.1903.06376}{arXiv:1903.06376[gr-qc]}
	
	\bibitem{Visinelli2019}
	C. Bambi, K. Freese, S. Vagnozzi and L. Visinelli,
	Testing the rotational nature of the supermassive object M87* from the circularity and size of its first image,
	\href{https://doi.org/10.1103/PhysRevD.100.044057}{Phys. Rev. D \textbf{100}, 044057 (2019).}
	\href{https://doi.org/10.48550/arXiv.1904.12983}{arXiv:1904.12983[gr-qc]}
	
	\bibitem{Konoplya2019}
	R. A. Konoplya, Shadow of a black hole surrounded by dark matter,
	\href{https://doi.org/10.1016/j.physletb.2019.05.043}{Phys. Lett. B \textbf{795}, 1-6 (2019).}
	\href{https://doi.org/10.48550/arXiv.1905.00064}{arXiv:1905.00064[gr-qc]}
	
	\bibitem{Gralla2019}
    S. E. Gralla and A. Lupsasca, Lensing by Kerr black holes, \href{https://doi.org/10.1103/PhysRevD.101.044031}{Phys. Rev. D \textbf{101}, 044031 (2020).}
    \href{https://doi.org/10.48550/arXiv.1910.12873}{arXiv:1910.12873[gr-qc]}
    
    \bibitem{LiuHS2019}
    H. S. Liu, Z. F. Mai, Y. Z. Li and H. L\"u, Quasi-topological Electromagnetism: Dark Energy, Dyonic Black Holes, Stable Photon Spheres and Hidden Electromagnetic Duality,
    \href{https://doi.org/10.1007/s11433-019-1446-1}{Sci. China Phys. Mech. Astron. \textbf{63}, 240411 (2020).}
    \href{https://doi.org/10.48550/arXiv.1907.10876}{arXiv:1907.10876[hep-th]}
    
    \bibitem{ZhuQH2020}
    Z. Chang and Q. H. Zhu, Revisiting a rotating black hole shadow with astrometric observables,
    \href{https://doi.org/10.1103/PhysRevD.101.084029}{Phys. Rev. D \textbf{101}, 084029 (2020).}
    \href{https://doi.org/10.48550/arXiv.2001.05175}{arXiv:2001.05175[gr-qc]}
    
    \bibitem{GuoM2020}
    M. Guo and P. -C. Li, Innermost stable circular orbit and shadow of the 4D Einstein–Gauss–Bonnet black hole, 
    \href{https://doi.org/10.1140/epjc/s10052-020-8164-7}{Eur. Phys. J. C {\bf 80}, 588 (2020).}  \href{https://arxiv.org/abs/2003.02523}{arXiv:2003.02523[gr-qc]}
	
	\bibitem{ZengXX2020}
	X. -X. Zeng, H. -Q. Zhang, H. -B. Zhang, Shadows and photon spheres with spherical accretions in the four-dimensional Gauss–Bonnet black hole, \href{https://doi.org/10.1140/epjc/s10052-020-08449-y}{Eur. Phys. J. C \textbf{80}, 872 (2020).}
	\href{https://doi.org/10.48550/arXiv.2004.12074}{arXiv:2004.12074[gr-qc]}
	
	\bibitem{Joshi2020}
	A. B. Joshi, D. Dey, P. S. Joshi and P. Bambhaniya, Shadow of a naked singularity without photon sphere, \href{https://doi.org/10.1103/PhysRevD.102.024022}{Phys. Rev. D \textbf{102}, 024022 (2020).}   \href{https://doi.org/10.48550/arXiv.2004.06525}{arXiv:2004.06525[gr-qc]}
	
	\bibitem{Lima2021}
	H. C. D. Lima, Junior., L. C. B. Crispino, P. V. P. Cunha and C. A. R. Herdeiro,
	Can different black holes cast the same shadow?,
	\href{https://doi.org/10.1103/PhysRevD.103.084040}{Phys. Rev. D \textbf{103}, 084040 (2021).}
	\href{https://doi.org/10.48550/arXiv.2102.07034}{arXiv:2102.07034[gr-qc]}
	
	\bibitem{GanQY2021a}	
	Q. Gan, P. Wang, H. Wu and H. Yang, Photon spheres and spherical accretion image of a hairy black hole, 
	\href{https://doi.org/10.1103/PhysRevD.104.024003}{Phys. Rev. D {\bf 104}, 024003 (2021).}   
	\href{https://arxiv.org/abs/2104.08703}{arXiv:2104.08703[gr-qc]}
	
	\bibitem{GanQY2021b}
	Q. Gan, P. Wang, H. Wu and H. Yang, Photon ring and observational appearance of a hairy black hole,
	\href{https://doi.org/10.1103/PhysRevD.104.044049}{Phys. Rev. D \textbf{104}, 044049 (2021).}
	\href{https://doi.org/10.48550/arXiv.2105.11770}{arXiv:2105.11770[gr-qc]}
	
	\bibitem{GaoSJ2022}
	M. Guo, Z. Zhong, J. Wang and S. Gao, Light rings and long-lived modes in quasiblack hole spacetimes,
	\href{https://doi.org/PhysRevD.105.024049}{Phys. Rev. D \textbf{105}, 024049 (2022).}
	\href{https://doi.org/10.48550/arXiv.2108.08967}{arXiv:2108.08967[gr-qc]}
	
	\bibitem{Adler2022}
	S. L. Adler and K. S. Virbhadra, Cosmological constant corrections to the photon sphere and black hole shadow radii,
	\href{https://doi.org/10.1007/s10714-022-02976-7}{Gen. Rel. Grav. \textbf{54}, 93 (2022).}
	\href{https://doi.org/10.48550/arXiv.2205.04628}{arXiv:2205.04628[gr-qc]}
	
	\bibitem{Perlick2022a}
	V. Perlick and O. Y. Tsupko, Calculating black hole shadows: Review of analytical studies, \href{https://doi.org/10.1016/j.physrep.2021.10.004}{Phys. Rept. {\bf 947}, 1–39 (2022).}   \href{https://doi.org/10.48550/arXiv.2105.07101}{arXiv:2105.07101[gr-qc]}
	
	\bibitem{WangMZ2022}
	M. Wang, S. Chen and J. Jing, Chaotic shadows of black holes: a short review, 
	\href{https://doi.org/10.1088/1572-9494/ac6e5c}{Commun. Theor. Phys. {\bf 74}, 097401 (2022).} \href{https://doi.org/10.48550/arXiv.2205.05855}{arXiv:2205.05855[gr-qc]}
	
	\bibitem{Vagnozzi2023}
	S. Vagnozzi, R. Roy, Y. D. Tsai, L. Visinelli, M. Afrin, A. Allahyari, P. Bambhaniya, D. Dey, S. G. Ghosh and P. S. Joshi, \emph{et al.}
	Horizon-scale tests of gravity theories and fundamental physics from the Event Horizon Telescope image of Sagittarius A,
	\href{https://doi.org/10.1088/1361-6382/acd97b}{Class. Quant. Grav. \textbf{40}, 165007 (2023).}
	\href{https://doi.org/10.48550/arXiv.2205.07787}{arXiv:2205.07787[gr-qc]}
	
	\bibitem{GuoGZ2023}
	G. Guo, Y. Lu, P. Wang, H. Wu and H. Yang, Black holes with multiple photon spheres,
	\href{https://doi.org/10.1103/PhysRevD.107.124037}{Phys. Rev. D \textbf{107}, 124037 (2023).}
	\href{https://doi.org/10.48550/arXiv.2212.12901}{arXiv:2212.12901[gr-qc]}
	
	\bibitem{Chen2023}
	S. Chen, J. Jing, W. -L. Qian and B. Wang, Black hole images: A Review, \href{https://doi.org/10.1007/s11433-022-2059-5}{Sci. China Phys. Mech. Astron. {\bf 66}, 260401 (2023).}
	\href{https://doi.org/10.48550/arXiv.2301.00113}{arXiv:2301.00113[astro-ph.HE]}
	
	\bibitem{Bargueno2023}
	P. Bargue\~no, Light rings in static and extremal black holes,
	\href{https://doi.org/10.1103/PhysRevD.107.104029}{Phys. Rev. D \textbf{107}, 104029 (2023).}
	\href{https://doi.org/10.48550/arXiv.2211.16899}{arXiv:2211.16899[gr-qc]}
	
	\bibitem{Ghosh2023}
	R. Ghosh, S. Sk and S. Sarkar, Hairy black holes: Nonexistence of short hairs and a bound on the light ring size,
	\href{https://doi.org/10.1103/PhysRevD.108.L041501}{Phys. Rev. D \textbf{108}, L041501 (2023).}
	\href{https://doi.org/10.48550/arXiv.2306.14193}{arXiv:2306.14193[gr-qc]}
	
	\bibitem{Tsukamoto2024}
	N. Tsukamoto,
	Circular light orbits of a general, static, and spherical symmetrical wormhole with $Z_2$ symmetry,
	\href{https://doi.org/10.1140/epjc/s10052-024-13696-4}{Eur. Phys. J. C \textbf{84}, 1325 (2024).}
	\href{https://doi.org/10.48550/arXiv.2401.07846}{arXiv:2401.07846[gr-qc].}
	
	\bibitem{Vertogradov2024}
	V. Vertogradov and A. \"Ovg\"un,
	Analyzing the influence of geometrical deformation on photon sphere and shadow radius: A new analytical approach \textemdash{} Spherically symmetric spacetimes,
	\href{https://doi.org/10.1016/j.dark.2024.101541}{Phys. Dark Univ. \textbf{45}, 101541 (2024).}
	\href{https://doi.org/10.48550/arXiv.2404.04046}{arXiv:2404.04046[gr-qc]}
	
	\bibitem{Murk2024}
	S. Murk and I. Soranidis, Light rings and causality for nonsingular ultracompact objects sourced by nonlinear electrodynamics,
	\href{https://doi.org/10.1103/PhysRevD.110.044064}{Phys. Rev. D \textbf{110}, 044064 (2024).} \href{https://doi.org/10.48550/arXiv.2406.07957}{arXiv:2406.07957[gr-qc].}
	
	\bibitem{Cederbaum2016}
	C. Cederbaum and G. J. Galloway, Uniqueness of photon spheres in electro-vacuum spacetimes,
	\href{https://doi.org/10.1088/0264-9381/33/7/075006}{Class. Quant. Grav. \textbf{33}, 075006 (2016).}
	\href{https://doi.org/10.48550/arXiv.1508.00355}{arXiv:1508.00355[math.DG]}
	
	\bibitem{Gibbons2016}
	M. Cvetic, G. W. Gibbons and C. N. Pope, Photon spheres and sonic horizons in black holes from supergravity and other theories, \href{https://doi.org/10.1103/PhysRevD.94.106005}{Phys. Rev. D \textbf{94}, 106005 (2016).}   \href{https://doi.org/10.48550/arXiv.1608.02202}{arXiv:1608.02202[gr-qc]}
	
	\bibitem{Berry2020}
	T. Berry, A. Simpson and M. Visser, Photon spheres, ISCOs, and OSCOs: Astrophysical observables for regular black holes with asymptotically Minkowski cores, \href{https://doi.org/10.3390/universe7010002}{Universe \textbf{7}, 2 (2020).}
	\href{https://doi.org/10.48550/arXiv.2008.13308}{arXiv:2008.13308[gr-qc]}
	
	\bibitem{Wielgus2021}
	M. Wielgus, Photon rings of spherically symmetric black holes and robust tests of non-Kerr metrics, 
	\href{https://doi.org/10.1103/PhysRevD.104.124058}{Phys. Rev. D \textbf{104}, 124058 (2021).}
	\href{https://doi.org/10.48550/arXiv.2109.10840}{arXiv:2109.10840[gr-qc]}
	
	\bibitem{Isomura2023}
	K. Isomura, R. Suzuki and S. Tomizawa, Particle motions around regular black holes,
	\href{https://doi.org/10.1103/PhysRevD.107.084003}{Phys. Rev. D \textbf{107}, 084003 (2023).}
	\href{https://doi.org/10.48550/arXiv.2301.10465}{arXiv:2301.10465[gr-qc]}
	
	\bibitem{Cunha2017}
	P. V. P. Cunha, E. Berti and C. A. R. Herdeiro, Light-Ring Stability for Ultracompact Objects,
	\href{https://doi.org/10.1103/PhysRevLett.119.251102}{Phys. Rev. Lett. {\bf 119}, 251102 (2017).}
	
	\bibitem{Cunha2020}
	P. V. P. Cunha and C. A. R. Herdeiro, Stationary Black Holes and Light Rings,
	\href{https://doi.org/10.1103/PhysRevLett.124.181101}{Phys. Rev. Lett. {\bf 124}, 181101 (2020).}
	\href{https://doi.org/10.48550/arXiv.2003.06445}{arXiv:2003.06445[gr-qc]}
	
	\bibitem{Hod2011}
	S. Hod, Hairy Black Holes and Null Circular Geodesics, \href{https://doi.org/10.1103/PhysRevD.84.124030}{Phys. Rev. D \textbf{84}, 124030 (2011)}   \href{https://doi.org/10.48550/arXiv.1112.3286}{arXiv:1112.3286[gr-qc]}
	
	\bibitem{Carroll}
	S. M. Carroll, \emph{Spacetime and Geometry: An Introduction to General Relativity}, (Cambridge University Press, Cambridge, 2019). 
	
	\bibitem{Hartle}
	J. B. Hartle, \emph{Gravity: An Introduction to Einstein's General Relativity}, (Cambridge University Press, Cambridge, 2021). 
	
	\bibitem{Straumann}
	N. Straumann, \emph{General Relativity} (Graduate Texts in Physics, Springer, Dordrecht, 2013).
	
	\bibitem{Raffaelli2021}	
	B. Raffaelli, Hidden conformal symmetry on the black hole photon sphere,   
	\href{https://doi.org/10.1007/JHEP03(2022)125}{J. High Energ. Phys. {\bf 2022}, 125 (2022).}
	\href{https://doi.org/10.48550/arXiv.2112.12543}{arXiv:2112.12543[gr-qc]}
	
	\bibitem{Destounis2023}
	K. Destounis, F. Angeloni, M. Vaglio and P. Pani,
	Extreme-mass-ratio inspirals into rotating boson stars: Nonintegrability, chaos, and transient resonances,
	\href{https://arxiv.org/10.1103/PhysRevD.108.084062}{Phys. Rev. D \textbf{108}, 8 (2023).}
	\href{https://doi.org/10.48550/arXiv.2305.05691}{arXiv:2305.05691[gr-qc]}
	
	\bibitem{Gibbons1993}
	G. W. Gibbons, No glory in cosmic string theory, \href{https://doi.org/10.1016/0370-2693(93)91278-U}{Phys. Lett. B {\bf 308}, 237-239 (1993).}
	
	\bibitem{Virbhadra2001}
	Clarissa-Marie Claudel, K. S. Virbhadra and G. F. R. Ellis, The geometry of photon surfaces,
	\href{https://doi.org/10.1063/1.1308507}{J. Math. Phys. {\bf 42}, 818-838 (2001).}
	\href{https://doi.org/10.48550/arXiv.gr-qc/0005050}{arXiv:0005050[gr-qc]}
	
	\bibitem{Koga2019}
	Y. Koga and T. Harada, Stability of null orbits on photon spheres and photon surfaces, \href{https://doi.org/10.1103/PhysRevD.100.064040}{Phys. Rev. D \textbf{100}, 064040 (2019).}   \href{https://doi.org/10.48550/arXiv.1907.07336}{arXiv:1907.07336[gr-qc]}
	
	\bibitem{Kobialko2022}
	K. Kobialko, I. Bogush and D. Gal'tsov, Geometry of massive particle surfaces,
	\href{https://doi.org/10.1103/PhysRevD.106.084032}{Phys. Rev. D \textbf{106}, 084032 (2022).}
	\href{https://doi.org/10.48550/arXiv.2208.02690}{arXiv:2208.02690[gr-qc]}
	
	\bibitem{SongY2023}
	Y. Song, Y. Cen, L. Tang, J. Hu, K. Diao, X. Zhao and S. Shi,
	The particle surface of spinning test particles,
	\href{https://doi.org/10.1140/epjc/s10052-023-11970-5}{Eur. Phys. J. C \textbf{83}, 833 (2023).}
	\href{https://doi.org/10.48550/arXiv.2208.03665}{arXiv:2208.03665[gr-qc]}
	
	\bibitem{Cunha2018}
	P. V. P. Cunha and C. A. R. Herdeiro, Shadows and strong gravitational lensing: a brief review, \href{https://doi.org/10.1007/s10714-018-2361-9}{Gen Relativ Gravit {\bf 50}, 42 (2018).}
	\href{https://doi.org/10.48550/arXiv.1801.00860}{arXiv:1801.00860[gr-qc]}
	
	\bibitem{WeiSW2020}
	S. -W. Wei, Topological charge and black hole photon spheres, \href{https://doi.org/10.48550/arXiv.2006.02112}{Phys. Rev. D \textbf{102}, 064039 (2020).}   \href{https://doi.org/10.48550/arXiv.2006.02112}{arXiv:2006.02112[gr-qc]}
	
	\bibitem{Lima2022}
	H. C. D. Lima Junior, J. -Z. Yang, L. C. B. Crispino, P. V. P. Cunha and C. A. R. Herdeiro,
	Einstein-Maxwell-dilaton neutral black holes in strong magnetic fields: Topological charge, shadows, and lensing,
	\href{https://doi.org/10.1103/PhysRevD.105.064070}{Phys. Rev. D \textbf{105}, 064070 (2022).}
	\href{https://doi.org/10.48550/arXiv.2112.10802}{arXiv:2112.10802[gr-qc]}
	
	\bibitem{WeiSW2023}
	S. -W. Wei and Y. -X. Liu, Topology of equatorial timelike circular orbits around stationary black holes, \href{https://doi.org/10.1103/PhysRevD.107.064006}{Phys. Rev. D {\bf 107}, 064006 (2023).}
	\href{https://doi.org/10.48550/arXiv.2207.08397}{arXiv:2207.08397[gr-qc]}
	
	\bibitem{WeiSW2023b}
	X. Ye and S. -W. Wei, Distinct topological configurations of equatorial timelike circular orbit for spherically symmetric (hairy) black holes, \href{https://doi.org/10.1088/1475-7516/2023/07/049}{J. Cosmol. Astropart. Phys. \textbf{2023(07)}, 049 (2023).}
	\href{https://doi.org/10.48550/arXiv.2301.04786}{arXiv:2301.04786[gr-qc]}
	
	\bibitem{WeiSW2023c}
	S. W. Wei, Y. P. Zhang, Y. X. Liu and R. B. Mann, Static spheres around spherically symmetric black hole spacetime,
	\href{https://doi.org/10.1103/PhysRevResearch.5.043050}{Phys. Rev. Res. \textbf{5}, 043050 (2023).}
	\href{https://doi.org/10.48550/arXiv.2303.06814}{arXiv:2303.06814[gr-qc]}
	
	\bibitem{YinJ2023}
	J. Yin, J. Jiang and M. Zhang, Kinematic topologies of black holes,
	\href{https://doi.org/10.1103/PhysRevD.108.044077}{Phys. Rev. D \textbf{108}, 044077 (2023).}
	\href{https://doi.org/10.48550/arXiv.2305.14179}{arXiv:2305.14179[gr-qc]}
	
	\bibitem{Tavlayan2023}
	A. Tavlayan and B. Tekin,
	Light rings around five dimensional stationary black holes and naked singularities,
	\href{https://doi.org/10.1103/PhysRevD.107.024016}{Phys. Rev. D \textbf{107}, 024016 (2023).}
	\href{https://doi.org/10.48550/arXiv.2209.14873}{arXiv:2209.14873[gr-qc]}
	
	\bibitem{Sadeghi2024}
	J. Sadeghi, M. A. S. Afshar, S. N. Gashti and M. R. Alipour, Thermodynamic topology and photon spheres in the hyperscaling violating black holes, \href{https://doi.org/10.1016/j.astropartphys.2023.102920}{Astropart.Phys. \textbf{156}  102920 (2024).}   \href{https://doi.org/10.48550/arXiv.2307.12873}{arXiv:2307.12873[gr-qc]}
	
	\bibitem{Cunha2024}
	P. V. P. Cunha, C. A. R. Herdeiro and J. P. A. Novo, Light rings on stationary axisymmetric spacetimes: Blind to the topology and able to coexist,
	\href{https://doi.org/10.1103/PhysRevD.109.064050}{Phys. Rev. D \textbf{109}, 064050 (2024).}
	\href{https://doi.org/10.48550/arXiv.2401.05495}{arXiv:2401.05495[gr-qc]}
	
	\bibitem{Xavier2024}
	S. V. M. C. B. Xavier, C. A. R. Herdeiro and L. C.B.Crispino, Traversable wormholes and light rings,
	\href{https://doi.org/10.1103/PhysRevD.109.124065}{Phys. Rev. D \textbf{109}, 124065 (2024).}
	\href{https://doi.org/10.48550/arXiv.2404.02208}{arXiv:2404.02208[gr-qc]}
	
	\bibitem{Afshar2024}
	J. Sadeghi and M. A. S. Afshar,
	The role of topological photon spheres in constraining the parameters of black holes,
	\href{https://doi.org/10.1016/j.astropartphys.2024.102994}{Astropart. Phys. \textbf{162}, 102994 (2024).}
    \href{https://doi.org/10.48550/arXiv.2405.06568}{arXiv:2405.06568[gr-qc]}
	
	\bibitem{Afshar2024b}
	J. Sadeghi and M. A. S. Afshar,
	Effective Potential and Topological Photon Spheres: A Novel Approach to Black Hole Parameter Classification,
	\href{https://doi.org/10.48550/arXiv.2405.18798}{arXiv:2405.18798[gr-qc]}
	
	\bibitem{WeiSW2022b}
	S. -W. Wei and Y. -X. Liu, Topology of black hole thermodynamics, \href{https://doi.org/10.1103/PhysRevD.105.104003}{Phys. Rev. D {\bf 105}, 104003 (2022).}  \href{https://doi.org/10.48550/arXiv.2112.01706}{arXiv:2112.01706[gr-qc]}
	
	\bibitem{WeiSW2022}
	S. -W. Wei, Y. -X. Liu and R. B. Mann, Black Hole Solutions as Topological Thermodynamic Defects, \href{https://doi.org/10.1103/PhysRevLett.129.191101}{Phys. Rev. Lett. {\bf 129}, 191101 (2022).}   \href{https://doi.org/10.48550/arXiv.2208.01932}{arXiv:2208.01932[gr-qc]}
	
    \bibitem{Qiao2022a}
    C.-K. Qiao and M. Li, A Geometric Approach on Circular Photon Orbits and Black Hole Shadow, \href{https://doi.org/10.1103/PhysRevD.106.L021501}{Phys. Rev. D {\bf 106}, L021501 (2022).} \href{https://doi.org/10.48550/arXiv.2204.07297}{arXiv:2204.07297[qr-qc]}
    
    \bibitem{Qiao2022b}
    C.-K. Qiao, Curvatures, photon spheres, and black hole shadows, \href{https://doi.org/10.1103/PhysRevD.106.084060}{Phys.Rev.D {\bf 106}, 084060 (2022).}   \href{https://doi.org/10.48550/arXiv.2208.01771}{arXiv:2208.01771[gr-qc]}
	
	\bibitem{Abramowicz1988}
	M. A. Abramowicz, B. Carter and J. P. Lasota, Optical reference geometry for stationary and static dynamics, 
	\href{https://doi.org/10.1007/BF00758937}{Gen. Relativ. Gravit. {\bf 20}, 1173–1183 (1988).}
	
	\bibitem{Gibbons2008}
	G. W. Gibbons and M. C. Werner, Applications of the Gauss-Bonnet theorem to gravitational lensing, 
	\href{https://doi.org/10.1088/0264-9381/25/23/235009}{Classical Quantum Gravity {\bf 25}, 235009 (2008).}  \href{https://arxiv.org/abs/0807.0854}{arXiv:0807.0854[gr-qc]}
	
	\bibitem{Gibbons2009}
	G. W. Gibbons and C. M. Warnick, Universal properties of the near-horizon optical geometry, 
	\href{https://doi.org/10.1103/PhysRevD.79.064031}{Phys. Rev. D {\bf 79}, 064031 (2009).}  \href{https://arxiv.org/abs/0809.1571}{arXiv:0809.1571[gr-qc]}
	
	\bibitem{Werner2012}	
	M. C. Werner, Gravitational lensing in the Kerr-Randers optical geometry, 
	\href{https://doi.org/10.1007/s10714-012-1458-9}{Gen. Relativ. Gravit. {\bf 44}, 3047-3057 (2012).}  \href{https://arxiv.org/abs/1205.3876}{arXiv:1205.3876[gr-qc]}
	
	\bibitem{Cunha2022}
	P. V. P. Cunha, C. A. R. Herdeiro and J. P. A. Novo, Null and timelike circular orbits from equivalent 2D metrics, 
	\href{https://doi.org/10.1088/1361-6382/ac987e}{Classical Quantum Gravity {\bf 39}, 225007 (2022).} \href{https://doi.org/10.48550/arXiv.2207.14506}{arXiv:2207.14506[gr-qc]}
	
	\bibitem{GaoSJ2021}
	M. Guo and S. Gao, Universal Properties of Light Rings for Stationary Axisymmetric Spacetimes,
	\href{https://doi.org/10.1103/PhysRevD.103.104031}{Phys. Rev. D \textbf{103}, 104031 (2021).}
	\href{https://doi.org/10.48550/arXiv.2011.02211}{arXiv:2011.02211[gr-qc]}
	
	\bibitem{Ghosh2021}
	R. Ghosh and S. Sarkar, Light rings of stationary spacetimes,
	\href{https://doi.org/10.1103/PhysRevD.104.044019}{Phys. Rev. D {\bf 104}, 044019 (2021).}
	\href{https://doi.org/10.48550/arXiv.2107.07370}{arXiv:2107.07370[gr-qc]} 
	
	\bibitem{Ishihara2016a}	
	A. Ishihara, Y. Suzuki, T. Ono, T. Kitamura, and H. Asada, Gravitational bending angle of light for finite distance and the Gauss-Bonnet theorem, \href{https://doi.org/10.1103/PhysRevD.94.084015}{Phys. Rev. D {\bf 94}, 084015 (2016).} \href{https://arxiv.org/abs/1604.08308}{arXiv:1604.08308[gr-qc]}
	
	\bibitem{Ishihara2016b}	
	A. Ishihara, Y. Suzuki, T. Ono and H. Asada, Finite-distance corrections to the gravitational bending angle of light in the strong deflection limit, \href{https://doi.org/10.1103/PhysRevD.95.044017}{Phys. Rev. D {\bf 95}, 044017 (2017).}  \href{https://arxiv.org/abs/1612.04044}{arXiv:1612.04044[gr-qc]}
	
	\bibitem{Ono2017}
	T. Ono, A. Ishihara and H. Asada, Gravitomagnetic bending angle of light with finite-distance corrections in stationary axisymmetric spacetimes, 
	\href{https://doi.org/10.1103/PhysRevD.96.104037}{Phys. Rev. D {\bf 96}, 104037 (2017).} \href{https://arxiv.org/abs/1704.05615}{arXiv:1704.05615[gr-qc]}
	
	\bibitem{Jusufi2018}		
	K. Jusufi, A. \"Ovg\"un, J. Saavedra, Y. V\'asquez and P. A. Gonz\'alez, Deflection of light by rotating regular black holes using the Gauss-Bonnet theorem, 
	\href{https://doi.org/10.1103/PhysRevD.97.124024}{Phys. Rev. D {\bf 97}, 124024 (2018).}  \href{https://arxiv.org/abs/1804.00643}{arXiv:1804.00643[gr-qc]}
	
	\bibitem{Jusufi2018b}
	K. Jusufi and A. \"Ovg\"un, \emph{Gravitational lensing by rotating wormholes}, 
	\href{https://doi.org/10.1103/PhysRevD.97.024042}{Phys. Rev. D {\bf 97}, 024042 (2018).}   \href{https://arxiv.org/abs/1708.06725}{arXiv:1708.06725[gr-qc]}
	
	\bibitem{Crisnejo2018}
	G. Crisnejo and E. Gallo, Weak lensing in a plasma medium and gravitational deflection of massive particles using the Gauss-Bonnet theorem. A unified treatment, \href{https://doi.org/10.1103/PhysRevD.97.124016}{Phys. Rev D {\bf 97}, 124016, (2018).}
	\href{https://doi.org/10.48550/arXiv.1804.05473}{arXiv:1804.05473[gr-qc]}
	
	\bibitem{Ono2019}
	T. Ono and H. Asada, The effects of finite distance on the gravitational deflection angle of light,
	\href{https://doi.org/10.3390/universe5110218}{Universe {\bf 5}, 218 (2019).}
	\href{https://doi.org/10.48550/arXiv.1906.02414}{arXiv:1906.02414[gr-qc]}
	
	\bibitem{Takizawa2020}	
	K. Takizawa, T. Ono, and H. Asada, Gravitational deflection angle of light: Definition by an observer and its application to an asymptotically nonflat spacetime, \href{https://doi.org/10.1103/PhysRevD.101.104032}{Phys. Rev. D {\bf 101}, 104032 (2020).}  \href{https://arxiv.org/abs/2001.03290}{arXiv:2001.03290[gr-qc]}
	
	\bibitem{LiZH2020}
	Z. Li and T. Zhou, Equivalence of Gibbons-Werner method to geodesics method in the study of gravitational lensing,
	\href{https://doi.org/10.1103/PhysRevD.101.044043}{Phys. Rev. D {\bf 101}, 044043 (2020).}
	\href{https://doi.org/10.48550/arXiv.1908.05592}{arXiv:1908.05592[gr-qc]}
	
	\bibitem{LiZH2020a}	
	Z. Li, J. Jia, The finite-distance gravitational deflection of massive particles in stationary spacetime: a Jacobi metric approach, \href{https://doi.org/10.1140/epjc/s10052-020-7665-8}{Eur. Phys. J. C {\bf 80}, 157 (2020).}  \href{https://arxiv.org/abs/1912.05194}{arXiv:1912.05194[gr-qc]}
	
	\bibitem{LiZH2020b}	
	Z. Li, G. Zhang and A. \"Ovg\"un, Circular orbit of a particle and weak gravitational lensing, 
	\href{https://doi.org/10.1103/PhysRevD.101.124058}{Phys. Rev. D {\bf 101}, 124058 (2020).}  \href{https://arxiv.org/abs/2006.13047}{arXiv:2006.13047[gr-qc]}
	
	\bibitem{Huang2022}
    Y. Huang and Z. Cao, Generalized Gibbons-Werner method for deflection angle, \href{https://doi.org/10.1103/PhysRevD.106.104043}{Phys. Rev. D {\bf 106}, 104043 (2022).}

    \bibitem{Huang2023}
    Y. Huang, Z. Cao and Z. Lu, Generalized Gibbons-Werner method for stationary spacetimes, \href{https://doi.org/10.1088/1475-7516/2024/01/013}{JCAP {\bf 2024(01)}, 013 (2024).}   \href{https://doi.org/10.48550/arXiv.2306.04145}{arXiv:2306.04145[gr-qc]}
    
    \bibitem{Takizawa2023}
    K. Takizawa and H. Asada, Gravitational lens on a static optical constant-curvature background: Its application to the Weyl gravity model,
    \href{https://doi.org/10.1103/PhysRevD.108.104055}{Phys. Rev. D {\bf 108}, 104055 (2023).}
    \href{https://doi.org/10.48550/arXiv.2304.02219}{arXiv:2304.02219[gr-qc]}
	
	\bibitem{LiZH2024a}
	Z. Li, A Novel Method for Calculating Deflection Angle, 
	\href{https://doi.org/10.48550/arXiv.2401.01532}{arXiv:2401.12525[gr-qc].}
	
	\bibitem{LiZH2024b}
	Z. Li, Gravitational Lensing Using Werner's Method in Cartesian-like Coordinates,
	\href{https://doi.org/10.48550/arXiv.2404.19658}{arXiv:2404.19658[gr-qc].}
	
    \bibitem{Perlick2000}
    V. Perlick, Ray Optics, Fermat’s Principle, and Applications to General Relativity, Springer, Berlin (2000).
    
    \bibitem{LinQ2008}
    X.-H. Ye and Q. Lin, Gravitational lensing analysed by the graded refractive index of a vacuum, \href{https://doi.org/10.1088/1464-4258/10/7/075001}{J. Opt. A: Pure Appl. Opt. {\bf 10}, 075001 (2008).}   \href{https://doi.org/10.48550/arXiv.0711.0633}{arXiv:0711.0633[gr-qc]}
    
    \bibitem{ChernSS}
    D. Bao, S.-S. Chern, Z. Shen, \emph{An Introduction to Riemann-Finsler Geometry}, Graduate Texts in Mathematics (GTM, volume 200), Springer, New York (2000). 
    
    \bibitem{Berger}
    M. Berger, \emph{A Panoramic View of Riemannian Geometry}, Springer-Verlag, Berlin (2003).  
    
    \bibitem{Berger2}	
    M. Berger and B. Gostiaux, \emph{Differential Geometry: Manifolds, Curves, and Surfaces}, Springer-Verlag, New York (1988). 
    
    \bibitem{Carmo1976}
    M. Do Carmo, \emph{Differential Geometry of Curves and Surfaces}, Prentice-Hall (1976). 
    
    \bibitem{ChernWH}
    W. -H. Chern, \emph{Differential Geometry}, Peking University Press, Beijing (2006). 
	
	\bibitem{footnote-Fermat}
	The classical Fermat’s principle states that: light rays always follow specific spatial curves such that the optical distance $s_{ab}^{\text{OP}}=\int_{a}^{b} n(x)dx$ (where $n(x)$ is the reflective index) or the light propagation time $t$ is minimized. However, the classical Fermat’s principle is restricted to flat spaces. Fortunately, this principle can be generalized to static or stationary curved spacetimes. In such spacetimes, a global choice of the time coordinate $t$ and always be made, and massless photons (which travel along null geodesics in spacetime geometry) starting from a fixed emission point at a given time $t_{a}$ would eventually make the arrival time $t_{b}$ minimal, satisfying the variational condition $\delta\big[\int_{a}^{b}dt\big]=0$. Therefore, the photon orbits must be spatial geodesics in the optical geometry. In this optical geometry, its metric $dt^{2}=g_{ij}^{\text{OP}}dx^{i}dx^{j}$ gives infinitesimal change of the stationary time square, and the spatial geodesics in optical geometry always satisfy the variation $\delta\big[\int_{a}^{b}dt\big]=0$. In the generalization of Fermat's principle to static or stationary curved spacetimes, the spatial length in optical geometry $l_{ab}^{OP}=\int_{a}^{b}dt=\int_{a}^{b}\sqrt{g_{ij}^{OP}dx^{i}dx^{j}}$ effectively plays the role of ``optical distance''. More discussion of the generalized Fermat's principle in a static or stationary curved spacetime can be found in references \cite{Werner2012,Jusufi2018,Gibbons2009,Perlick2000,LinQ2008}.
    
    \bibitem{footnote notation}
    Mathematically, in the 2-dimensional curved surface, each continuous curve $\gamma=\gamma(s)$ can be assigned with a geodesic curvature $\kappa_{g}(\gamma)$. Here, we use the notation $\kappa_{g}(r=r_{ph})$ or $\kappa_{g}(r)$ to shown that we only focus on the geodesic curvature of circular curves with a constant radius $r$.
    
    \bibitem{footnote continuous}
    The geodesic curvature, as expressed explicitly in equation (\ref{geodesic cuurvature expression}) (or equation (\ref{geodesic-Curvature1})), depends on the first-order derivatives of the metric components in optical geometry (as well as the first-order derivatives of the metric components in 4-dimensional spacetime geometry), which must be continuous. The continuity of the first-order derivatives of metric components is always required to give the well-defined concepts of curvature scalar and curvature tensors, both in spacetime geometry and in optical geometry.
    
    \bibitem{footnote Gauss-Bonnet discontinuity}
    Conventionally, the Gauss-Bonnet theorem is applied to a smooth region, where the Gaussian curvature is continuous in the entire region of $D$ (and the geodesic curvature along boundary $\partial D$ is also continuous). The discontinuity of curvatures may potentially bring up additional terms contributing to the Gauss-Bonnet theorem (which may depend on the specific behaviors of the discontinuous points). However, exploring the possible extended formulas for the Gauss-Bonnet theorem in cases with discontinuities is far beyond the scope of our work. Our analysis is carried out assuming that the conventional Gauss-Bonnet theorem in equation (\ref{Gauss-Bonnet theorem}) is still valid. 
    
    \bibitem{footnote-geodesic-complete}
    Although a black hole spacetime has singularities, its optical geometry is usually defined outside the event horizon (where spacetime singularities are escaped from the cosmic censorship conjecture). Therefore, the optical geometry can be generally considered as a geodesic complete manifold, satisfying the prerequisite of Cartan-Hadamard theorem. Specifically, for static or stationary spacetimes, the time coordinate $t$ in spacetime geometry becomes an arc length parameter in optical geometry, the endless and infinitely extendibility of the observer's stationary time coordinate $t$ outside event horizons implies the geodesic completeness of optical geometry.
	
	
\end{thebibliography}

\clearpage

\begin{widetext}
	
\section*{Supplemental Material}

\begin{center}
{Chen-Kai Qiao}
\\
{College of Science, Chongqing University of Technology, Banan, Chongqing, 400054, China}
\end{center}

For the readers who are unfamiliar with the relation between geodesic curvature $\kappa_{g}$ principal curvatures $\mathcal{K}_{1}$, $\mathcal{K}_{2}$ and Gaussian curvature $\mathcal{K}$, we provide a concise introduction here. 

\begin{figure}[b]
	\includegraphics[width=0.625\textwidth]{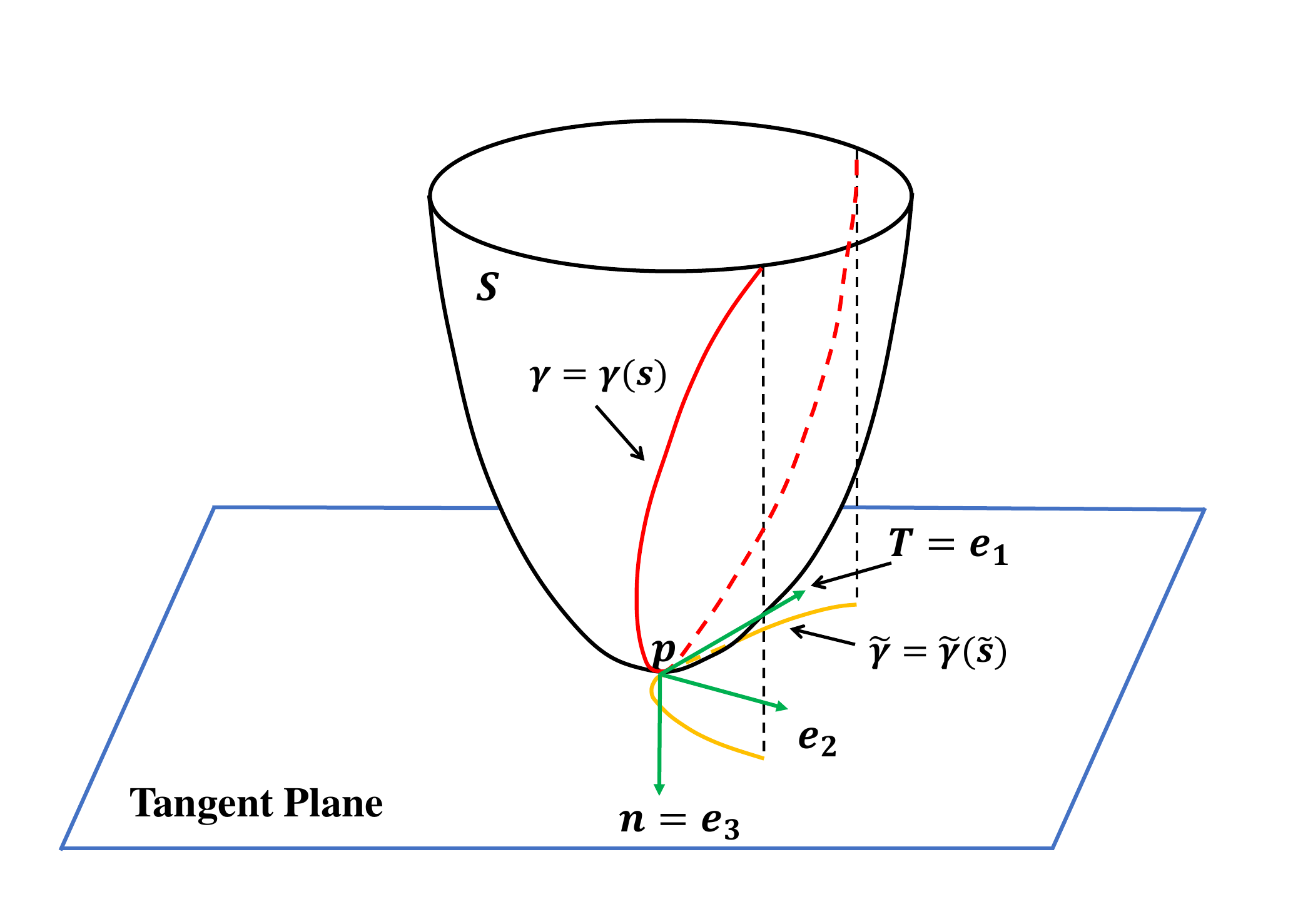}
	\caption{Illustration of the embedding of surface $S$ in a 3-dimensional Euclidean space. In this figure, the $\gamma=\gamma(s)$ is a continuous curve residing on surface $S$, and $p$ is an arbitrary point along this curve. The set $\{p; \boldsymbol{e_{1}},\boldsymbol{e_{2}},\boldsymbol{e_{3}}\}$ corresponds to frame fields in 3-dimensional Euclidean space such that $\boldsymbol{e_{1}}=\boldsymbol{T}$ is the unit tangent vector of curve $\gamma=\gamma(s)$ and $\boldsymbol{e_{3}}=\boldsymbol{n}$ is the unit normal vector of surface $S$ at point $p$. The tangent plane of curved surface $S$ at point $p$ are spanned by frames $\boldsymbol{e_{1}}$ and $\boldsymbol{e_{2}}$. The curve $\widetilde{\gamma}=\widetilde{\gamma}(\widetilde{s})$ represents the projection of $\gamma=\gamma(s)$ onto the tangent plane of surface $S$ at point $p$.}
	\label{figure Embedding}
\end{figure}

In the classical surface theory and differential geometry, we can assign geodesic curvature and normal curvature to a continuous curve $\gamma=\gamma(s)$ on a 2-dimensional surface $S$. The geodesic curvature $\kappa_{g}$ is an intrinsic quantity of the 2-dimensional surface $S$, while the normal curvature $\kappa_{n}$ relies on the embedding of surface $S$ into higher dimensional space. Considering the surface $S$ embedded into a 3-dimensional Euclidean space, which is illustrated in figure \ref{figure Embedding}, we can assign the frame fields $\{p; \boldsymbol{e_{1}},\boldsymbol{e_{2}},\boldsymbol{e_{3}}\}$ in the 3-dimensional background Euclidean space. The $\boldsymbol{e_{1}}=\boldsymbol{T}$ represents the unit tangent vector of curve $\gamma=\gamma(s)$, and $\boldsymbol{e_{3}}=\boldsymbol{n}$ denotes the unit normal vector of surface $S$ at point $p$. The tangent plane of curved surface $S$ at point $p$ is spanned by frames $\boldsymbol{e_{1}}$ and $\boldsymbol{e_{2}}$. The geodesic curvature $\kappa_{g}$ and normal curvature $\kappa_{n}$ of a continuous curve $\gamma=\gamma(s)$ are defined as: 
\begin{subequations}
\begin{eqnarray}
	\kappa_{g} & = & \frac{d\boldsymbol{T}}{ds} \cdot \boldsymbol{e_{2}} 
	= \frac{d^{2}\boldsymbol{r}(s)}{ds^{2}} \cdot \boldsymbol{e_{2}} \ , 
	\\
	\kappa_{n} & = & \frac{d\boldsymbol{T}}{ds} \cdot \boldsymbol{n} 
	= \frac{d^{2}\boldsymbol{r}(s)}{ds^{2}} \cdot \boldsymbol{e_{3}} \ .
\end{eqnarray}
\end{subequations}
Using geodesic curvature and normal curvature, the total curvature of a curve $\gamma(s)$ in 3-dimensional Euclidean space can be expressed by the square root $\kappa=\sqrt{\kappa_{g}^{2}+\kappa_{n}^{2}}$. In particular, curves with zero geodesic curvature ($\kappa_{g}=0$) are intrinsically flat on surface $S$, and they are referred to geodesics on this surface. Curves with zero total curvature ($\kappa=\sqrt{\kappa_{g}^{2}+\kappa_{n}^{2}}=0$) are precisely straight lines (geodesics) in 3-dimensional Euclidean space. The normal curvature $\kappa_{n}$ is the extrinsic quantity of surface $S$, which also depends on the embedding of surface into a higher dimensional space. Particularly, for a congruence of continuous curves on surface $S$ at a given point $p$, the changing of tangent direction of a curve $\gamma(s)$ (in the congruence of curves) at point $p$ would result in the variation of normal curvature $\kappa_{n}$ (notably, the congruence of curves at point $p$ sharing the same tangent direction have the same normal curvature). In certain circumstances, the normal curvature $\kappa_{n}$ could reach its maximum and minimum values. These extreme values reflect the geometric properties of a 2-dimensional surface $S$, and they are named as the principal curvatures of this surface at point $p$
\begin{equation}
	\mathcal{K}_{1} = \text{max}\{\kappa_{n}\} 
	\ \ \ \ \ 
	\mathcal{K}_{2} = \text{min}\{\kappa_{n}\} \ .
\end{equation}
The Gaussian curvature of surface $S$ (at point $p$) is defined to be the product of principal curvatures
\begin{equation}
	\mathcal{K} \equiv \mathcal{K}_{1} \cdot \mathcal{K}_{2} \ .
\end{equation}
The two principal curvatures $\mathcal{K}_{1}$ and $\mathcal{K}_{2}$ rely on the embedding of the 2-dimensional surface $S$ into a higher dimensional space / spacetime. However, their product (the Gaussian curvature $\mathcal{K} = \mathcal{K}_{1} \cdot \mathcal{K}_{2}$) is an intrinsic geometric quantity of this surfece $S$, regardless of the embedding. Mathematically, this fundamental result is given by the renowned theorem in surface theory --- the Gauss Theorema Egregium. Table \ref{table2} provides a summary of the characteristics associated with geodesic curvature $\kappa_{g}$, total curvature $\kappa$ and Gaussian curvature $\mathcal{K}$.

\begin{table*}
	\caption{This table summarizes the characteristics of geodesic curvature $\kappa_{g}$, total curvature $\kappa$ and Gaussian curvature $\mathcal{K}$, as well as their differences.}
	\label{table2}
	\vspace{2mm}
	\begin{ruledtabular}
		\begin{tabular}{lllll}
			& Geodesic Curvature & Total Curvature & Gaussian Curvature 
			\\
			\hline
			Notation & $\kappa_{g}$ & $\kappa=\sqrt{\kappa_{g}^{2}+\kappa_{n}^{2}}$ & $\mathcal{K} = \mathcal{K}_{1} \cdot \mathcal{K}_{2}$
			\\
			\hline
			Meaning & Curvature of a curve $\gamma$ (viewed  & Curvature of a curve $\gamma$ (embedded  & Intrinsic curvature of a two 
			\\
			& on the 2-dimensional surface $S$) & into higher dimensional space &  dimensional surface $S$
			\\
			\hline
			Property & It measures how much this curve  & It measures how much this curve  & It measures how much this two 
			\\
			& departs from being a geodesic & departs from being a geodesic &  dimensional surface $S$ deviates   
			\\
			& curve on 2-dimensional surface $S$. \ & curve in higher dimensional space. \ \  & from being a flat space intrinsically. 
			\\
			\hline
			Intrinsic/Extrinsic & $\kappa_{g}$ is intrinsic quantity of $S$ & $\kappa$ is extrinsic quantity of $S$ & $\mathcal{K}$ is intrinsic quantity of $S$ &
			\\
			& (It is independent of the & (It depends on the embedding & (It is independent of the 
			\\
			&  embedding of surface $S$ into & of surface $S$ into a higher &  embedding of surface $S$ into 
			\\
			& a higher dimensional space.) & dimensional space.) &  a higher dimensional space.)
			\\
			\hline
			Zero Curvature & $\kappa_{g}=0 \ \Leftrightarrow\ $ $\gamma$ is a geodesic curve & $\kappa=0 \ \Leftrightarrow\ $ $\gamma$ is a geodesic curve & $\mathcal{K}=0 \ \Leftrightarrow\ $ the 2-dimensional
			\\
			Condition & on the 2-dimensional surface $S$ & in higher dimensional space &  surface $S$ is intrinsically flat \footnote{However, the surface $S$ may exhibit nonzero “extrinsic curvature” when embedded into a higher dimensional space.}
		\end{tabular}
	\end{ruledtabular}
\end{table*}

Furthermore, in spacetime geometry (which is a four dimensional Lorentz manifold), one can select a two dimensional surface $S_{\mu\nu}$ using two tangent vector fields $\partial_{\mu}$ and $\partial_{\nu}$. For each of these two dimensional surfaces, the Gaussian curvature and geodesic curvature can be defined in similar ways. Particularly, the definition of Gaussian curvature is similar to the expression (\ref{Gauss-Curvature1}) in the maintext, namely via $\mathcal{K}=\frac{R_{\mu\nu\mu\nu}}{g_{\mu\mu}g_{\nu\nu}-(g_{\mu\nu})^{2}}$. In this way, a collection of Gaussian curvatures can be obtained in a four dimensional Lorentz manifold. Mathematicians often name these Gaussian curvatures (for different two dimensional surfaces $S_{\mu\nu}$) as sectional curvatures of a Lorentz manifold.

\clearpage

\end{widetext}

\end{document}